\documentclass[aps,prb,reprint,amsmath,amssymb,groupedaddress]{revtex4-2}
\usepackage{graphicx}
\usepackage{hyperref}

\begin{document}

\title{Discovery of High-Voltage Magnesium-Ion Cathodes using Machine Learning and First-Principles Calculations}

\author{Jhon Rogelnor A. Florida}
\email{florida.ja81@s.msumain.edu.ph}
\affiliation{Department of Physics, Mindanao State University - Main Campus, Marawi City, Philippines}

\author{Edward Aris D. Fajardo}
\affiliation{Department of Physics, Mindanao State University - Main Campus, Marawi City, Philippines}

\date{\today}

\begin{abstract}
Developing high-performance cathode materials for magnesium-ion batteries (MIBs) remains challenging because Mg$^{2+}$ ions move slowly, and conventional materials exhibit low voltage outputs. In this study, machine learning and first-principles calculations were combined to investigate topological quantum materials (TQMs) as a new class of cathode candidates. A modified crystal graph convolutional neural network (mCGCNN) was used to screen 917 Mg-containing TQMs, identifying a small subset of materials with predicted voltages above 3 V and high volumetric capacities. Among these, Mg$_2$VO$_4$ and Mg$_6$MnO$_8$ were selected for detailed density functional theory (DFT) analysis. Formation energy and convex-hull calculations indicate that Mg$_x$VO$_4$ exhibits a fully stable magnesiation pathway, whereas Mg$_x$MnO$_8$ demonstrates minor metastability at intermediate compositions. The calculated voltage profiles yield average voltages of 3.66 V for Mg$_2$VO$_4$ and 4.06 V for Mg$_6$MnO$_8$, in good agreement with machine learning predictions. Electronic structure analysis, supported by Wannier interpolation, confirms that both materials are semiconducting, with valence bands dominated by O $2p$ states and conduction bands by transition-metal $d$ states, indicating a charge-transfer redox mechanism. Compared to conventional Mg cathodes, these TQMs exhibit higher voltages and competitive capacities, underscoring their potential for next-generation multivalent batteries. This study demonstrates that integrating machine learning with first-principles calculations offers an efficient approach for discovering and understanding novel cathode materials.
 
\end{abstract}


\maketitle

\section{\label{sec:level1}Introduction}

Magnesium-ion batteries (MIBs) have emerged as a promising alternative to lithium-ion systems for next-generation energy storage, driven by the demand for safer, cost-effective, and high-capacity technologies \cite{yoo2013mg, mohtadi2014magnesium}. Although lithium-ion batteries (LIBs) offer high energy density and long cycle life, their drawbacks—including dendrite formation, safety concerns, and limited lithium resources—drive the search for alternative chemistries \cite{chen2021review,adebanjo2025comprehensive}. Magnesium metal offers several intrinsic benefits, such as dendrite-free deposition, high volumetric capacity (3833 mAh cm${-3}$), and low cost \cite{yoo2013mg, yagi2014concept, mohtadi2014magnesium}. However, progress in MIBs is limited by strong electrostatic interactions between Mg$^{2+}$ ions and host lattices, causing sluggish diffusion kinetics, limited rate capability, and low operating voltages \cite{levi2009review, lapidus2014solvation, rajput2018elucidating, okoshi2013theoretical, wan2015mg, saha2014rechargeable}. Conventional cathode materials, including Chevrel phases, spinel oxides, and layered transition-metal oxides, have had only partial success, typically showing restricted voltage and capacity. For example, Chevrel-phase Mo$_6$S$_8$ delivers about 1.2 V and moderate capacity \cite{aurbach2000prototype}. Recent advances such as spinel MgMn$_2$O$_4$, NASICON-type NaV$_2$(PO$_4$)$_3$, and MgFeSiO$_4$ demonstrate improved voltage and structural properties, highlighting the importance of open diffusion pathways and transition-metal (TM) redox activity \cite{okamoto2015intercalation, zeng2017promising, orikasa2014high, kaewmaraya2014electronic, deng2025advanced, saranya2025comparative}. Still, developing cathode materials with voltages above 3 V and volumetric capacities over 800 mAh cm$^{-3}$ remains a major challenge \cite{chen2025ai}. Recently, topological quantum materials (TQMs), defined by symmetry-protected electronic states and high carrier mobility, have been proposed as a new class of functional materials with potential advantages in electronic transport and stability \cite{obeid2022recent, yi2019topological, wu2020coupling}. While TQMs have shown promise in lithium-, sodium-, and zinc-ion batteries \cite{liu2017all, yi2019topological, zhao2022few, wang2023screening, wu2021screening}, their use in magnesium-ion systems is largely unexplored. 

Integrating machine learning (ML) with first-principles calculations has become a powerful way to accelerate materials discovery. Graph-based neural network models (GNN), including CGCNN\cite{xie2018crystal}, ALIGNN\cite{choudhary2021atomistic}, MEGNet\cite{chen2019graph}, and SchNet\cite{schutt2018schnet}, predict material properties directly from crystal structures and have been effectively applied to battery materials research\cite{dick2020machine, zeng2023deepmd, gong2023general}. These methods use large datasets from materials databases such as the Materials Project (MP)\cite{jain2013commentary} and Inorganic Crystal Structure Database (ICSD)\cite{hellenbrandt2004inorganic}, but often focus on lithium-based systems, limiting their use for multivalent ions. Recent advances in modified CGCNN (mCGCNN) models have greatly improved predictive accuracy across various ion chemistries, achieving mean absolute errors of 0.25–0.33 V for voltage prediction \cite{chen2025ai, wang2024integrating, zhang2022interpretable}. ML screening has also been applied to topological materials, identifying candidates with high capacity and favorable transport properties \cite{wang2023screening}. These developments show that combining ML with DFT validation provides an efficient and physically grounded framework for discovering novel cathode materials.

This study employs a modified CGCNN model in inference mode to screen 917 magnesium-containing topological quantum materials from the topological quantum chemistry (TQC) database, focusing on candidates with voltages above 3 V and volumetric capacities greater than 800 mAh cm$^{-3}$ \cite{chen2025ai}. In Sec.~II, we describe the computational framework, including the machine learning screening and first-principles methods. In Sec.~III, we present the results and discussion. Specifically, we analyze the thermodynamic stability of the selected candidates through formation energy and convex-hull analysis, evaluate their electrochemical performance via voltage profiles, and examine their electronic structures using DFT and Wannier interpolation to elucidate the underlying redox mechanisms. Finally, Sec.~IV summarizes the main findings and conclusions of this work.

\section{Computational Methods}
Machine learning screening was performed to identify candidate materials with predicted voltages exceeding 3 V and volumetric capacities above 800 mAh cm$^{-3}$. Promising nontrivial candidates, Mg$_2$VO$_4$ and Mg$_6$MnO$_8$, were subsequently investigated using first-principles calculations.

\subsection{Data Collection}

A total of 917 Mg-containing topological quantum materials (TQMs) were collected from the topological quantum chemistry (TQC) database (\url{https://topologicalquantumchemistry.com/, https://www.cryst.ehu.es/})\cite{bradlyn2017topological,vergniory2019complete,vergniory2022all} and used as input to the trained modified crystal graph convolutional neural network (mCGCNN) model. The corresponding crystallographic information files (CIFs) were obtained from the Materials Project (MP) database \cite{jain2013commentary}.

\subsection{Modified Crystal Graph Convolutional Neural Network}
A modified CGCNN (mCGCNN) was used in inference mode to predict the voltage of TQMs directly from their crystal structures. In this framework, each crystal is described as a multigraph $G$ where atoms are treated as nodes and interatomic interactions within a specified cutoff distance are represented as edges. Structural information is learned through message passing between neighboring atoms across successive graph convolutional layers, following the crystal graph convolutional neural network architecture originally introduced for materials property prediction \cite{xie2018crystal}.Graph neural network models such as CGCNN have been widely applied for predicting battery-related properties, including electrode voltages and ionic conductivity, due to their ability to directly learn structural descriptors from crystal structures without the need for handcrafted features \cite{chen2025ai, zhang2022interpretable}. The initial atom feature vectors $v_i^{(0)}$ were adopted from the original CGCNN implementation. Interatomic distances between atoms $i$ and $j$ are encoded into edge feature vectors using a Gaussian basis expansion,
\begin{equation}
  u_{(i,j)_k}^{(t)} =
  \exp\!\left[
    -\left(
      \frac{d_{(i,j)_k}-\mu_t}{\sigma}
    \right)^2
  \right],
  \label{eq:gauss}
\end{equation}
where $d_{(i,j)_k}$ denotes the distance between atoms $i$ and $j$ at their $k$th edge, and
$\mu_t = 0.2t~\text{\AA}$ for $t = 0, 1, \ldots, 40$.
The atomic feature vectors of neighboring atoms and their corresponding edge features are then combined into a local environment descriptor,
\begin{equation}
  z_{(i,j)_k}^{(t)} = v_i^{(t)} \otimes v_j^{(t)} \otimes u_{(i,j)_k}
\end{equation}
which captures both chemical and structural information of the atomic environment. These descriptors are propagated through graph convolutional layers according to
\begin{align}
  v_i^{(t+1)}
  & = v_i^{(t)} + \sum_{j,k}
    \Big[
    \sigma\!\left(
    z_{(i,j)_k}^{(t)} W_f^{(t)} + b_f^{(t)}
    \right) \nonumber \\
  & \quad \quad \odot
    g\!\left(
    z_{(i,j)_k}^{(t)} W_s^{(t)} + b_s^{(t)}
    \right)
    \Big]
    \label{eq:cgcnn}
\end{align}
where $\odot$ denotes element-wise multiplication, $\sigma$ is the sigmoid function, and $g$ is a nonlinear activation function. $W$ and $b$ represent the trainable weights and biases of the network, respectively. After three convolutional layers, an average pooling operation is applied over the working ions to obtain a single feature vector $v_c$ that summarizes the cathode material. This vector is then passed through a fully connected layer to predict the average intercalation voltage of the crystal structure. This graph-based representation enables the model to capture the local chemical environment governing electrochemical behavior~\cite{chen2025ai}.

The predicted voltage is physically related to the change in Gibbs free energy during the intercalation reaction,
\begin{equation}
  E = -\frac{\Delta G}{nF}
  \label{eq:voltage}
\end{equation}
where E is the electrode voltage, $\Delta G$ is the Gibbs free energy change of the electrochemical reaction, $n$ is the number of transferred electrons, and $F$ is Faraday's constant~\cite{chen2025ai, urban2016}. Because the CGCNN framework aggregates structural and chemical information from neighboring atoms through graph convolutions, the model can effectively capture the key structural features governing the voltage of intercalation cathodes. Recent studies have demonstrated that graph neural network models can predict electrode voltages with mean absolute errors on the order of 0.25–0.33 V, making them effective tools for rapid screening of candidate battery materials~\cite{chen2025ai,zhang2022interpretable}.

Therefore, although the model was trained on a dataset of conventional intercalation-type battery materials, it can be applied as a screening tool for the 917 TQMs considered in this work. The graph-based representation allows the model to learn general relationships between crystal structure, local bonding environments, and electrochemical properties.

\subsection{First-principles Calculation of Promising High-voltage Materials}
The most promising cathode candidates are Mg$_2$VO$_4$ and Mg$_6$MnO$_8$,  screened out with mCGCNN. DFT calculations were performed using Quantum Espresso (QE) \cite{giannozzi2009quantum}. The crystallographic information files (CIFs) were obtained from the Materials Project database~\cite{jain2013commentary} and symmetric structure were generated using the pymatgen package. The calculations employed projected augmented wave (PAW) pseudopotentials and the exchange--correlation functional was described with generalized gradient approximation (GGA)\cite{perdew1992accurate} in Perdew-Burke-Ernzerhof (PBE) form\cite{perdew1996generalized}. A plane wave cut-off energy of 680 ev and a Monkhorst-Pack k-point mesh with a resolution of $0.18\,\mathrm{\AA^{-1}}$ ($7\times7\times7$ grid for the unit cell) is used for the necessary Brillion zone (BZ)\cite{monkhorst1976special} density. To account for the strong on-site Coulomb interaction of localized transition-metal $d$ electrons, a Hubbard $U$ correction was applied within the GGA+$U$ framework~\cite{zhou2004first,cococcioni2005linear, wang2006oxidation,jain2011formation}. The effective $U$ values were set to 3.25 eV for V and 5.0 eV for Mn, which are commonly adopted values for transition-metal oxides in previous computational studies~\cite{munro2020improved,ebrahimzadeh2025accelerated,wang2004enthalpy,basnet2022controlling}. The theoretical capacity was calculated using the formula
\begin{equation}
  C_{\mathrm{theor}} =  \frac{n F}{3.6M}
\end{equation}
where C represents the theoretical capacity mAh $\mathrm{g^{-1}}$, $n$ is the number of electrons exchange per formula unit , F is  Faraday's constant, and $M$ is the molar mass of the active material. To determine the reversible capacity and voltage profile, the relaxed crystal structure was first used as the starting point to generate intermediate compositions with different Mg concentrations by sequentially removing Mg atoms from the fully magnesiated structure. Since the Mg atoms occupy symmetry-equivalent or nearly equivalent sites, removing Mg from different positions results in nearly identical total energies. Therefore, one representative configuration was used for each composition by selecting Mg atoms to remove from the structure. The formation energy of each intermediate composition was calculated from the total energies of the optimized structures to construct the formation energy diagram as a function of Mg concentration. The convex hull was then determined to identify the thermodynamically stable intermediate phases during Mg extraction. In addition, the energy above the hull was evaluated for each composition, and structures with energies within 0.1 eV/atom above the hull were considered thermodynamically stable or metastable \cite{sun2016thermodynamic}.

For Mg-ion systems, each $\mathrm{Mg^{2+}}$ contributes two electrons, which can enhance the overall performance. The average voltage of a magnesium-ion cathode is calculated using
\begin{equation}
  V = \frac{E(\text{Mg}_{x_1}\text{MX}) -E(\text{Mg}_{x_2}\text{MX}) + (x_2 - x_1) E(\text{Mg})}{2(x_2 - x_1)e},
  \label{eq:mg_voltage}
\end{equation}
where $E(\text{Mg}_{x}\text{MX})$ is the total DFT energy of the cathode in the magnesium composition $x$, $E(\text{Mg})$ is the energy of metallic magnesium, and $e$ is the elementary charge \cite{han2018density,wang2023screening}.
 
\section{Results and Discussion}
\begin{figure}
	\centering
	\includegraphics[width=1\linewidth]{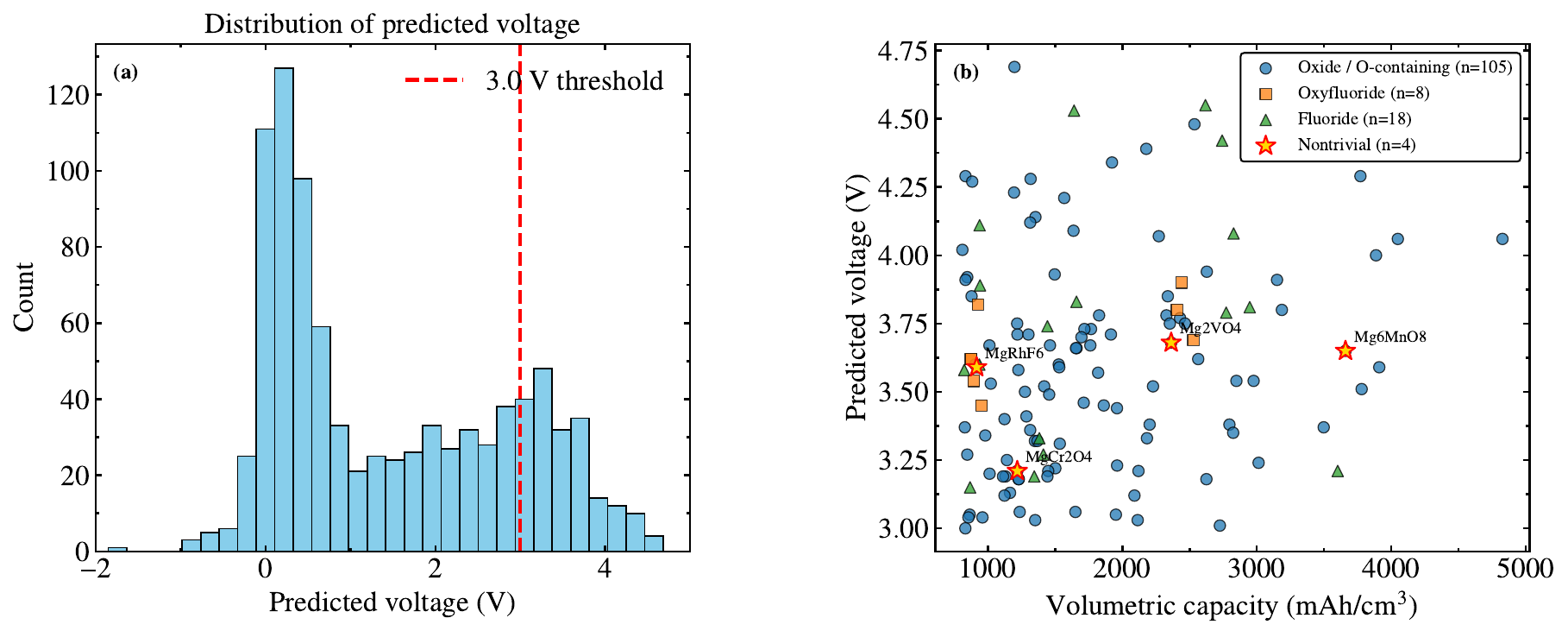}
	\caption{(a) Distribution of predicted voltages for 917 magnesium-containing topological quantum materials obtained from the machine learning screening. The dashed line indicates the 3.0 V threshold used to identify high-voltage candidates. (b) Voltage–capacity map of the screened materials, showing the relationship between predicted voltage and volumetric capacity. Oxide, fluoride, and oxyfluoride chemistries are distinguished, while nontrivial topological materials are highlighted as red stars.}
	\label{fig:histo}
\end{figure}

Machine learning screening was performed on 917 topological quantum materials to predict their intercalation voltages. The model has been shown to achieve reliable accuracy for voltage prediction in multivalent battery systems~\cite{chen2025ai}. The distribution of predicted voltages is shown in Fig.~\ref{fig:histo}(a). Most materials are concentrated in the low-voltage region (0–1 V), indicating that a large fraction of the dataset is unsuitable for high-voltage cathode applications. A smaller subset extends into the high-voltage region (2–4.5 V), with the 3 V threshold used to identify promising candidates.

\begin{table}[t]
  \caption{Predicted electrochemical properties of selected topological materials proposed as potential Mg-ion cathodes. MP-ID refers to the Materials Project identifier, and topology labels indicate semimetal (SM) or topological insulator (TI).}
  \label{tab:candidates}
  \renewcommand{\arraystretch}{1.25}
  \setlength{\tabcolsep}{3pt}
  \begin{tabular}{ccccc}
    \hline \hline
    MP-ID & Formula & $V$ (V) & $C_{\mathrm{vol}}$ (mAh cm$^{-3}$) & Topology \\
    \hline
    mp-30545 & Mg$_2$VO$_4$   & 3.68 & 2363 & SM \\
    mp-19239 & Mg$_6$MnO$_8$  & 3.65 & 3659 & SM \\
    mp-9927  & MgRhF$_6$      & 3.59 & 917  & SM \\
    mp-19202 & MgCr$_2$O$_4$  & 3.21 & 1219 & TI \\
    \hline \hline
  \end{tabular}
\end{table}

The voltage–capacity relationship is presented in Fig.~\ref{fig:histo}(b). Most materials cluster in a moderate-performance region with voltages between 3.0 and 4.2 V and volumetric capacities of 1000–3000 mAh cm$^{-3}$. Oxide-based compounds dominate this region, reflecting their structural stability and ability to accommodate Mg insertion. Fluoride systems tend to exhibit higher voltages but more scattered capacities, while oxyfluorides display intermediate behavior. Notably, the nontrivial materials (highlighted as red stars) occupy a relatively sparse region of the phase space. Among these, Mg$_6$MnO$_8$ combines high voltage with significantly larger volumetric capacity, while Mg$_2$VO$_4$ exhibits a balanced combination of voltage and capacity. All identified candidates exceed the 3 V threshold, motivating further first-principles investigation.

A complete list of candidate TQMs is presented in Table~\ref{tab:candidates}. Among the four identified compounds, Mg$_2$VO$_4$ and Mg$_6$MnO$_8$ exhibit relatively high volumetric capacities and were therefore selected for further investigation using first-principles calculations.

\subsection{Crystal Structure and Structural Optimization}

\begin{figure}[!h]
  \centering
  \includegraphics[width=1.1\linewidth]{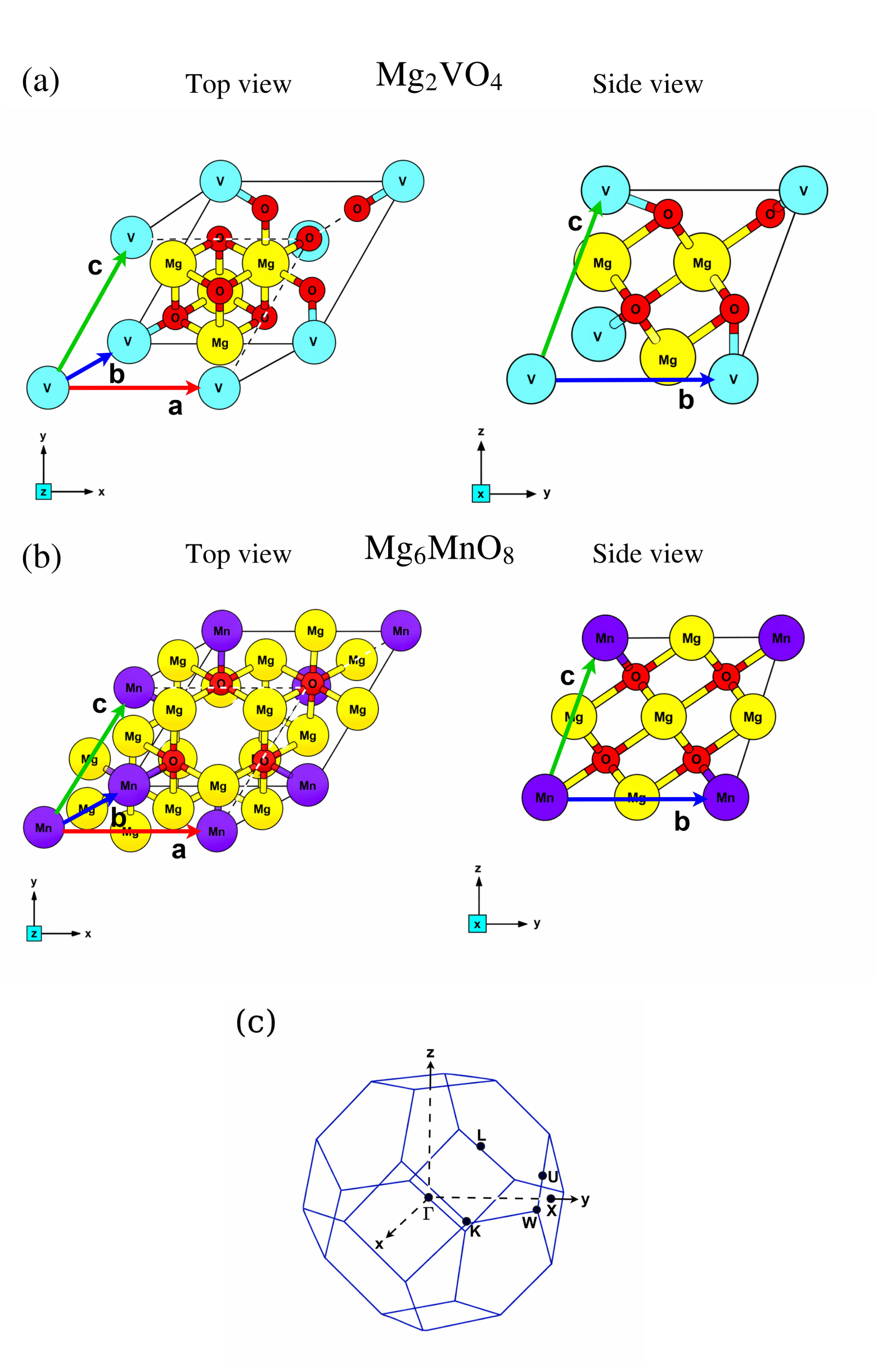}
  \caption{
    Crystal structures of (a) Mg$_2$VO$_4$ and (b) Mg$_6$MnO$_8$. Colors denote V atoms (cyan), Mn atoms (violet), Mg (yellow) atoms, and O atoms (red). Lattice vectors and Cartesian axes are indicated. (c) First Brillouin zone with the high-symmetry k-point path used for band structure calculations.
  }
  \label{fig:both}
\end{figure}

The Mg$_2$VO$_4$ and Mg$_6$MnO$_8$ crystallize in the cubic space group Fd$\bar{3}$m (No.~227). Structural relaxation was performed using variable-cell optimization, allowing both lattice parameters and atomic positions, relaxed to equilibrium, show in Fig.~\ref{fig:both}. In both systems, the optimized structures preserve the cubic symmetry, indicating structural stability upon relaxation. The optimized lattice parameters correspond to a primitive rhombohedral representation with $a=b=c\approx 5.9567$~\AA{} for Mg$_2$VO$_4$ and $a=b=c\approx 5.9823$~\AA{} for Mg$_6$MnO$_8$, with $\alpha=\beta=\gamma=60^\circ$, consistent with the primitive cell of an face-centered cubic (FCC) lattice. In Mg$_2$VO$_4$, V atoms occupy tetrahedrally coordinated sites forming VO$_4$ units, while Mg resides in octahedral environments. In contrast, Mg$_6$MnO$_8$ exhibits octahedral coordination of Mn (MnO$_6$), with Mg occupying interstitial sites within the oxygen framework. 

\subsection{Thermodynamic Stability for the Selected Candidates}

\begin{figure}
  \centering
  \includegraphics[width=1\linewidth]{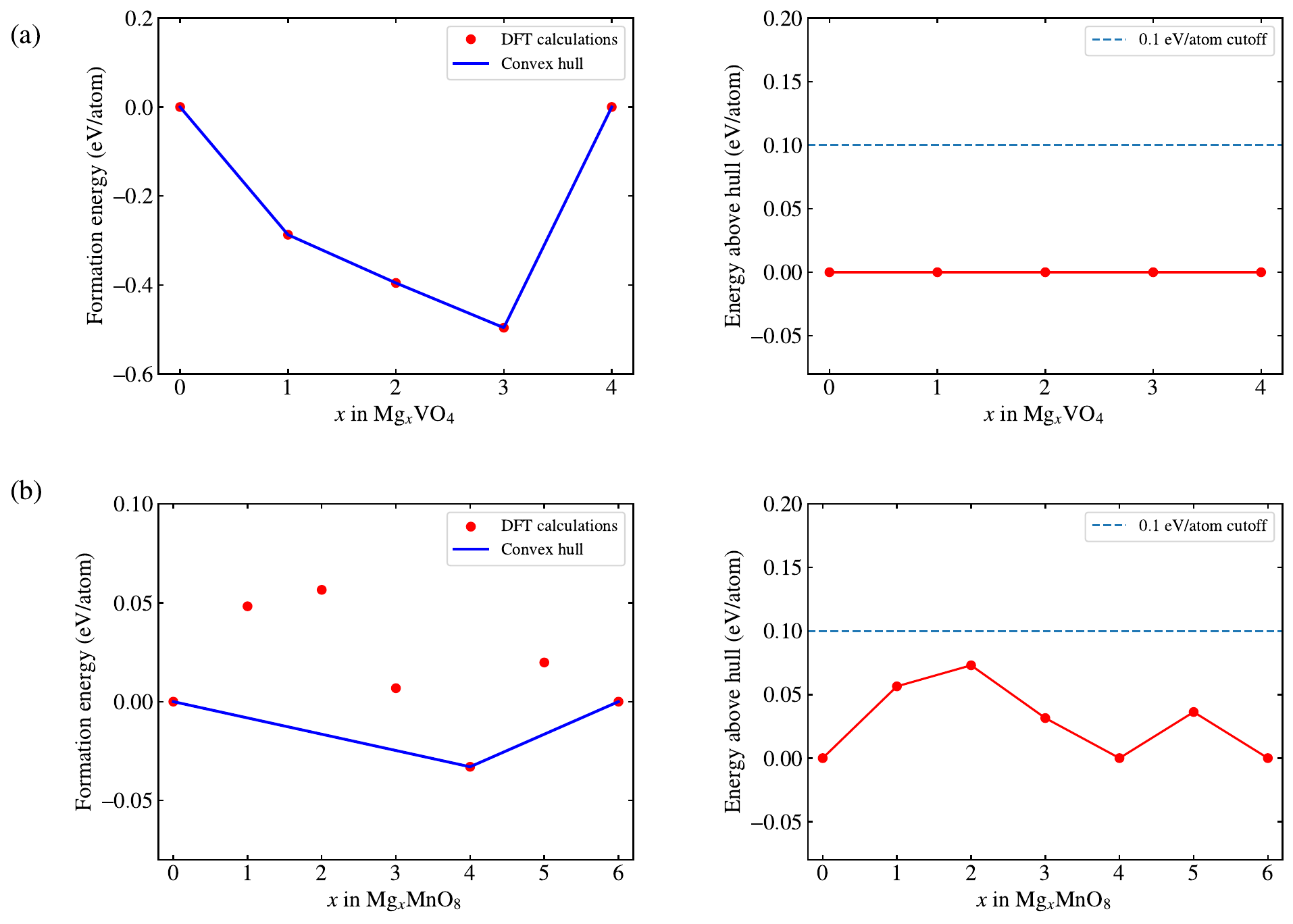}
  \caption{Formation energy and convex-hull analysis for (a) Mg$_x$VO$_4$ and (b) Mg$_x$MnO$_8$ as a function of Mg concentration. Solid lines represent the convex hull, while points indicate calculated intermediate compositions. Mg$_x$VO$_4$ lies entirely on the convex hull, indicating full thermodynamic stability, whereas Mg$_x$MnO$_8$ exhibits small deviations corresponding to metastable intermediate states.}
  \label{fig:form}
\end{figure}

The thermodynamic stability of Mg insertion in Mg$_x$VO$_4$ and Mg$_x$MnO$_8$ was evaluated through formation energy calculations and convex-hull analysis, as shown in Fig.~\ref{fig:form}. The formation energy describes the relative stability of intermediate Mg compositions with respect to the demagnesiated and fully magnesiated end members, while the energy above the convex hull provides a quantitative measure of metastability. For Mg$_x$VO$_4$, the formation energies remain negative across the entire Mg composition range, indicating that Mg insertion is energetically favorable at all concentrations as shown in Fig.~\ref{fig:form}(a). All intermediate structures lie directly on the convex hull, corresponding to zero energy above hull. This demonstrates that each composition represents a thermodynamically stable phase and that the system evolves through a sequence of equilibrium states during Mg insertion and extraction. Such behavior reflects a well-defined intercalation mechanism, in which structural rearrangements occur without destabilizing the host lattice. In contrast, Mg$_x$MnO$_8$ exhibits small deviations from the convex hull at several intermediate compositions as shown in Fig.~\ref{fig:form}(b). The calculated energies above hull range from approximately 0.03 to 0.07 eV/atom, indicating the presence of metastable configurations. Despite this, these values remain below the commonly accepted metastability threshold of $\sim$0.1 eV/atom~\cite{sun2016thermodynamic}, suggesting that these phases are likely to be accessible under practical electrochemical conditions. The deviations from the convex hull indicate that the system may locally favor phase separation; however, the relatively small magnitude of the instability implies that kinetic effects and finite-temperature contributions can stabilize these intermediate states during cycling. The difference in stability between the two systems reflects the underlying electronic and structural response to Mg insertion. In Mg$_x$VO$_4$, the smooth evolution of formation energies indicates a continuous redistribution of electronic charge within the V–O framework, which stabilizes all intermediate compositions. In contrast, Mg$_x$MnO$_8$ exhibits competing electronic configurations associated with multiple Mn oxidation states, leading to slight energetic imbalances between intermediate structures. As a result, Mg$_x$VO$_4$ follows a fully stable magnesiation pathway, while Mg$_x$MnO$_8$ exhibits minor metastability that remains within the range typically considered acceptable for battery materials. These differences in thermodynamic stability are directly reflected in the voltage profiles discussed in the following section.

\subsection{Electrochemical Voltage Profile}

\begin{figure}
  \centering
  \includegraphics[width=1\linewidth]{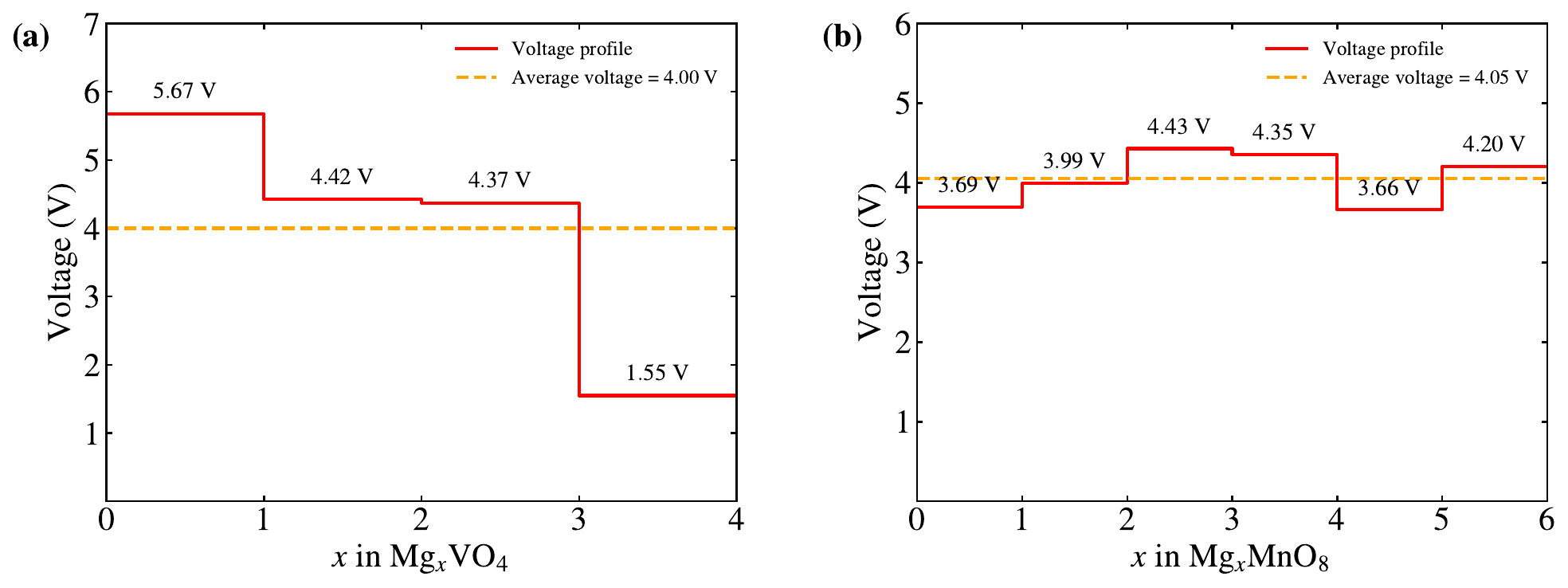}
  \caption{Calculated voltage profiles for (a) Mg$_x$VO$_4$ and (b) Mg$_x$MnO$_8$ as a function of Mg concentration. Mg$_x$VO$_4$ exhibits a high initial voltage followed by a sharp decrease at high Mg content, while Mg$_x$MnO$_8$ shows a relatively stable voltage plateau, reflecting differences in the underlying formation-energy landscape.}
  \label{fig:voltage}
\end{figure}

The voltage profiles of Mg$_x$VO$_4$ and Mg$_x$MnO$_8$ are shown in Fig.~\ref{fig:voltage}. For Mg$_x$VO$_4$, the voltage decreases from a high initial value of approximately 5.67 V to 4.42 V and 4.37 V, followed by a sharp drop to about 1.55 V at high Mg concentration. This behavior reflects a strong initial interaction between Mg ions and the host structure, which weakens as the structure becomes increasingly saturated. The sharp voltage drop at high Mg content indicates that the final insertion steps provide minimal additional stabilization.

In contrast, Mg$_x$MnO$_8$ exhibits a relatively flat voltage profile, with values ranging from approximately 3.66 V to 4.43 V. This indicates that the energy differences between successive Mg configurations are more uniform, leading to a stable voltage output throughout the insertion process. The difference between the two systems reflects the underlying formation-energy landscape: Mg$_x$VO$_4$ shows greater stability variations across intermediate states, while Mg$_x$MnO$_8$ exhibits more uniform energetics. These results demonstrate that the voltage profile is directly governed by the thermodynamic stability of intermediate phases.

The reversible capacities of Mg$_2$VO$_4$, Mg$_6$MnO$_8$, and MgCr$_2$O$_4$ are summarized in Table~\ref{tab:performance}. For Mg$_2$VO$_4$, a reversible capacity of 655 mAh g$^{-1}$ is obtained from the full Mg insertion range, consistent with its fully stable convex-hull behavior. Mg$_6$MnO$_8$ exhibits a higher reversible capacity of 978 mAh g$^{-1}$ due to its ability to accommodate a larger number of Mg ions. Although some intermediate states are metastable, their energies remain within the accessible range, enabling reversible Mg insertion. For comparison, MgCr$_2$O$_4$ shows a lower capacity of 280 mAh g$^{-1}$, as reported in previous first-principles studies~\cite{chen2018first}, where Mg-vacancy ordering limits reversible insertion.

\begin{table}[htbp]
  \caption{Performance comparison of selected candidate cathode materials for magnesium-ion batteries (MIBs).}
  \label{tab:performance}
  \setlength{\tabcolsep}{5pt}
  \begin{tabular}{lccc}
    \hline \hline
    Material & $C_\mathrm{rev}$ (mAh g$^{-1}$) & $V$ (V) & $E_\mathrm{dens}$ (Wh kg$^{-1}$) \\
    \hline
    Mg$_2$VO$_4$ & 655 & 4.00 & 2397.3 \\
    Mg$_6$MnO$_8$ & 978 & 4.05 & 3970.7 \\
    MgCr$_2$O$_4$ & 280 & 3.6 \footnote{Ref.~\onlinecite{chen2018first}} & 1008.0 \\
    \hline \hline
  \end{tabular}  
\end{table}

\begin{figure}
  \centering
  \includegraphics[width=1\linewidth]{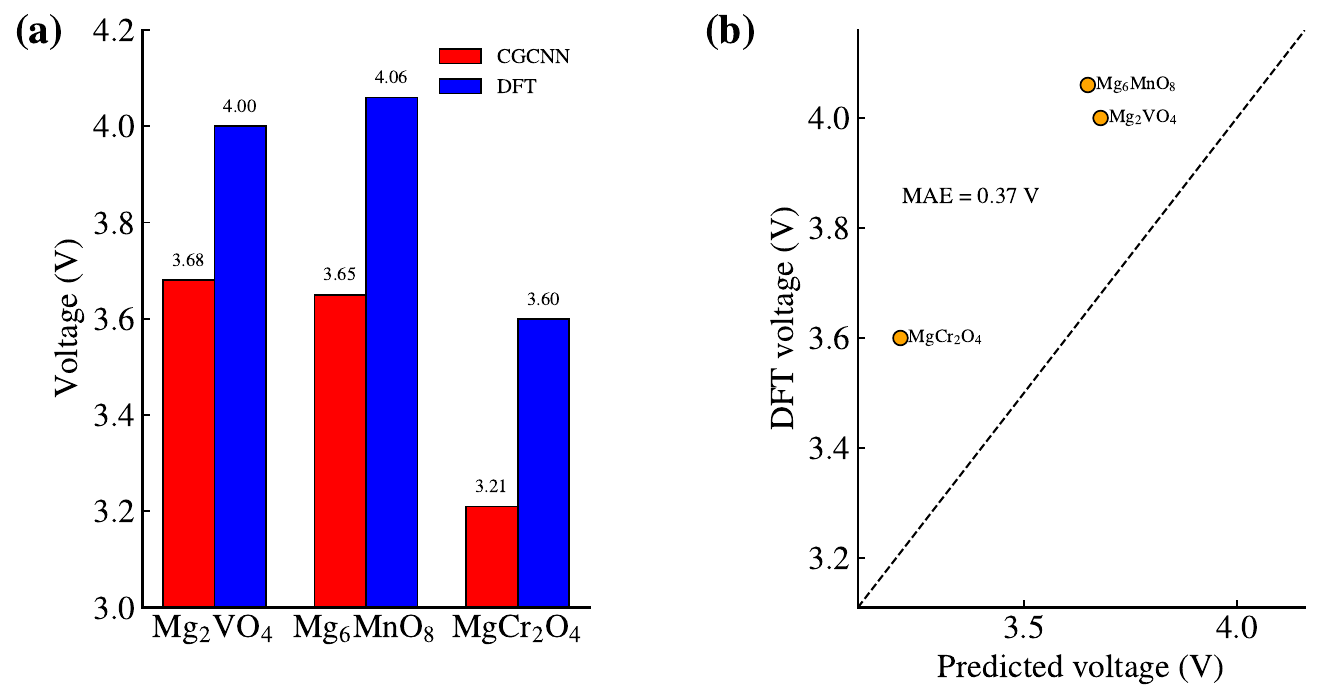}
  \caption{(a) Comparison of machine learning-predicted voltages and DFT-calculated values for selected Mg-ion cathode materials. (b) Parity plot showing the agreement between predicted and DFT voltages. The dashed line represents perfect agreement.}
  \label{fig:comp}
\end{figure}

Figure~\ref{fig:comp}(a) compares the voltages predicted by the modified CGCNN model with those obtained from DFT calculations. The model consistently underestimates the voltage for all considered materials. In the case of Mg$_2$VO$_4$, the predicted value (3.68 V) is lower than the DFT result (4.00 V), with similar deviations observed for Mg$_6$MnO$_8$ and MgCr$_2$O$_4$. This systematic trend is evident in the parity plot shown in Fig.~\ref{fig:comp}(b), where all data points lie above the ideal agreement line, yielding a mean absolute error of approximately 0.37 V. Despite this deviation, the model preserves the relative ordering of materials, indicating that it retains predictive value for screening purposes.

The observed discrepancy can be traced to the underlying design of the modified CGCNN model. By construction, the model assumes that the intercalation voltage is primarily governed by the local chemical environment of the working ions, which is encoded through graph convolution and pooling over Mg-centered features\cite{chen2025ai}. While this assumption is effective for conventional intercalation cathodes, the present results indicate that it is insufficient for fully describing the electrochemical behavior of the studied systems. In particular, the voltage is strongly influenced by transition metal redox activity and structural relaxation during Mg insertion, both of which involve non-local and electronic effects that are not explicitly captured by the model. These limitations become more pronounced when the model is applied to TQMs, which often exhibit more complex electronic structures. Consequently, although the model provides reasonable estimates for high-throughput screening, its quantitative accuracy is reduced when evaluated against first-principles calculations, underscoring the need for DFT validation in such systems.

\begin{figure*}[!t]
	\centering
	\includegraphics[width=1\linewidth]{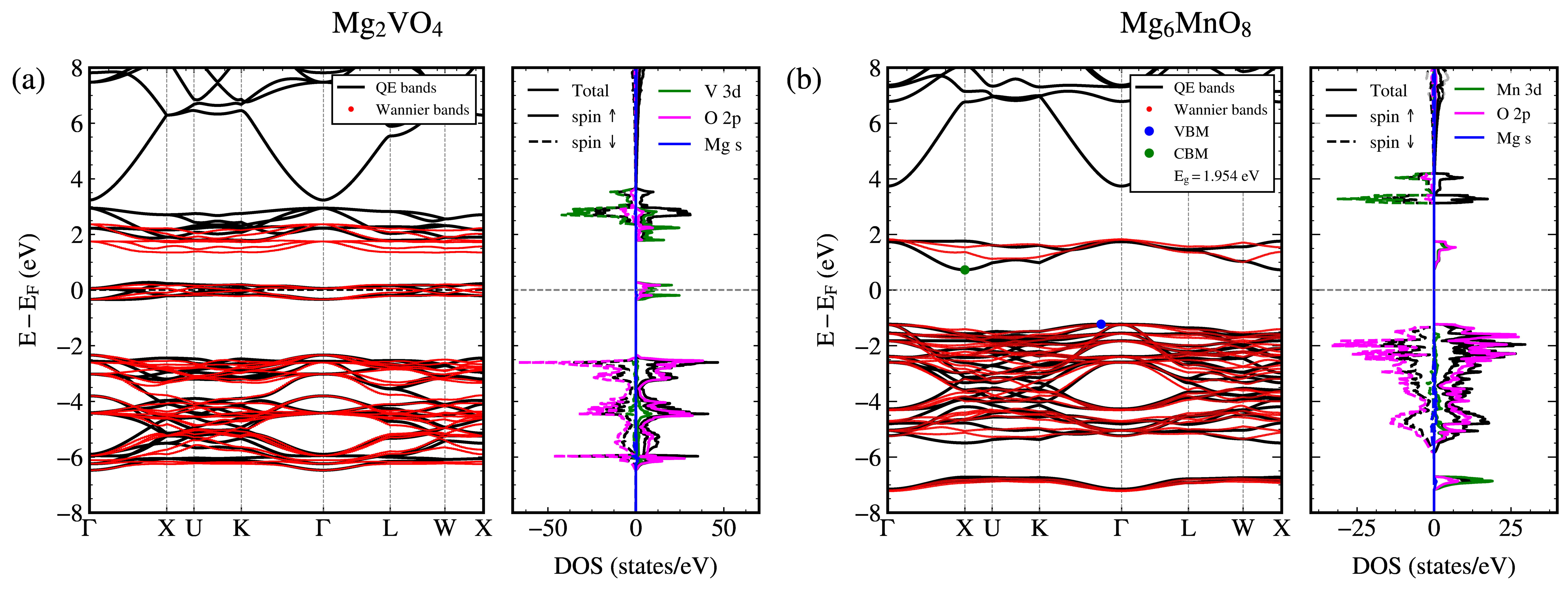}
	\caption{Electronic band structures and density of states (DOS) for (a) Mg$_2$VO$_4$ and (b) Mg$_6$MnO$_8$. Solid lines represent DFT results, while dashed lines correspond to Wannier-interpolated bands. The Fermi level is set to zero. Mg$_2$VO$_4$ shows semimetallic behavior with strong O $2p$–V $3d$ hybridization, whereas Mg$_6$MnO$_8$ exhibits a semiconducting gap with spin-polarized Mn $3d$ states.}
	\label{fig:band_dos}
\end{figure*}

\subsection{Electronic Structure: Band Dispersion and Orbital Contributions}
The electronic band structures of Mg$_2$VO$_4$ and  Mg$_6$MnO$_8$ were analyzed using DFT and further validated through Wannier interpolation. Figure~\ref{fig:both}(c) shows the corresponding Brillouin zone of the face-centered cubic lattice. As shown in Fig.~\ref{fig:band_dos}(a), the electronic band structure of Mg$_2$VO$_4$ lies at the threshold between semiconducting and metallic behavior, with bands located very close to the Fermi level. The bands near E$_F$are relatively flat, indicating limited dispersion and suggesting low carrier mobility. Comparison between the DFT and Wannier-interpolated bands reveals strong agreement, particularly in the deeper valence region where the bands are smooth and well reproduced. This demonstrates that the Wannier basis effectively captures the dominant orbital characteristics of the system. Minor deviations persist near the Fermi level, where the bands exhibit heightened sensitivity. Overall, the Wannier-interpolated bands closely overlap with the DFT bands throughout the Brillouin zone, confirming that the low-energy electronic structure is accurately represented by the constructed tight-binding Hamiltonian. Near the Fermi level, the bands display moderate dispersion, indicating partially delocalized electronic states capable of supporting efficient charge transport. The density of states (DOS) analysis reveals that these states are primarily of V $3d$ character, with significant hybridization from O $2p$ orbitals in the valence region. This O $2p$–V $3d$ hybridization suggests a charge-transfer-type electronic structure, in which electrons donated by Mg$^{2+}$ are accommodated within the V–O framework. Since the V $3d$ states are situated very close to the Fermi level, electron insertion can occur with minimal energy cost, accounting for both the high initial voltage and the rapid voltage decrease at elevated Mg concentrations as these states become occupied. The lack of pronounced spin splitting and the strong agreement between Wannier and DFT results further support a relatively simple and stable redox mechanism, consistent with the convex-hull stability observed across all Mg compositions. 

In contrast, Mg$_6$MnO$_8$ exhibits well-defined semiconducting behavior with a band gap of approximately 1.95 eV as shown in Fig.~\ref{fig:band_dos}(b). The separation between the valence-band maximum and the conduction-band minimum is clearly visible, indicating a more electronically insulating ground state than in the V-based system. The Wannier-interpolated bands closely reproduce the DFT band structure across the entire energy window, demonstrating that the electronic states near the Fermi level are also well described by a localized orbital basis. This confirms the reliability of the Wannier-derived tight-binding model for analyzing the system's low-energy electronic properties. The DOS analysis indicates that the valence band is dominated by O $2p$ states, while the conduction band primarily consists of Mn $3d$ orbitals exhibiting clear spin polarization. The observed spin splitting reflects strong electron correlation and magnetic interactions associated with Mn, signifying a more complex electronic environment compared to the V-based system. The conduction bands display moderate dispersion, suggesting that electron transport is feasible but more restricted than in Mg$_2$VO$_4$. The greater separation between O $2p$ and Mn $3d$ states results in a more stable electronic configuration during Mg insertion, consistent with the relatively uniform and elevated voltage profile. Nevertheless, the increased localization of Mn $3d$ states and the presence of multiple oxidation states introduce competing electronic configurations, which likely contribute to the small but finite energy above the convex hull observed in intermediate compositions.

\section{Conclusion}
In summary, both machine learning and first-principles calculations were employed to identify promising topological quantum materials (TQMs) for magnesium-ion battery (MIB) cathodes. A modified crystal graph convolutional neural network (mCGCNN) was used to screen 917 magnesium-containing TQMs based on predicted voltages and volumetric capacities. Only a limited subset of these materials satisfied the target criteria of a voltage above 3 V and a volumetric capacity exceeding 800 mAh $cm^{-3}$, highlighting the challenge of developing high-performance magnesium cathodes. From this subset, four notable candidates were identified, with Mg$_2$VO$_4$ and Mg$_6$MnO$_8$ selected for further investigation. Thermodynamic analysis of formation energies and convex-hull construction, with reversible capacities of 655 mAh $g^{-1}$ and 978 mAh $g^{-1}$ respectively, revealed distinct behaviors: Mg$_x$V$_2$O$_8$ follows a stable magnesiation pathway, with all intermediates located on the convex hull, indicating equilibrium phases during magnesium insertion. In contrast, Mg$_x$MnO$_8$ exhibits small energies above the hull for certain intermediate states, suggesting the presence of metastable phases; however, these remain within the commonly accepted stability limit of 0.1 eV/atom and are potentially accessible during operation. Comparison between predicted and DFT-calculated voltages shows a systematic underestimation by the model across all materials. For Mg$_2$VO$_4$, the predicted voltage of 3.68 V is lower than the DFT value of approximately 4.00 V, with similar trends observed for Mg$_6$MnO$_8$ and MgCr$_2$O$_4$. Despite this deviation, the model correctly identifies high-voltage candidates and preserves the relative performance trends, confirming its effectiveness for high-throughput screening. 
Electronic structure analysis using Wannier interpolation indicates that these materials are topological semiconductors, characterized by clear band gaps and the absence of states crossing the Fermi level. The valence bands are predominantly O $2p$ states, while the conduction bands are mainly transition-metal d states (V $3d$ or Mn $3d$), suggesting redox activity via charge transfer. Compared to conventional magnesium cathodes such as Chevrel-phase Mo$_6$S$_8$~\cite{aurbach2000prototype}, spinel MgMn$_2$O$_4$~\cite{okamoto2015intercalation, truong2017unravelling}, and layered V$_2$O$_5$~\cite{mukherjee2020, huie2015cathode}, the identified TQMs exhibit higher predicted voltages and comparable capacities. These materials may also offer enhanced electronic transport due to their unique electronic structures. Although challenges related to magnesium diffusion and electrolyte compatibility persist, the results indicate that TQMs represent promising and largely unexplored cathode materials for multivalent ion batteries.

\begin{acknowledgments}
We gratefully acknowledge the support of the Department of Science and Technology – Science Education Institute (DOST-SEI) through the STRAND Scholarship Program for providing assistance in completing this research study.

\end{acknowledgments}

\appendix
\section{Derivation of the Average Voltage for Mg-Ion Cathodes}%
\label{app:voltage}

The average intercalation voltage of a magnesium-ion cathode can be derived from thermodynamic considerations. In general, the cell voltage is related to the change in Gibbs free energy associated with Mg insertion according to
\begin{equation}
  V = -\frac{\Delta G}{z e},
\end{equation}
where $\Delta G$ is the Gibbs free energy change of the reaction, $z$ is the number of electrons transferred per Mg ion ($z = 2$ for Mg$^{2+}$), and $e$ is the elementary charge. This relationship forms the basis of first-principles voltage calculations in battery materials \cite{han2018density}.

The Gibbs free energy can be expressed as
\begin{equation}
  \Delta G = \Delta E + P \Delta V - T \Delta S,
\end{equation}
where $\Delta E$ is the internal energy obtained from density functional theory (DFT), $P \Delta V$ is the pressure–volume contribution, and $T \Delta S$ is the entropic term. For solid-state intercalation reactions, both $P \Delta V$ and $T \Delta S$ are typically small compared to the total energy differences and can be neglected, allowing the voltage to be approximated directly from DFT total energies \cite{han2018density}. Thus, the voltage can be written as
\begin{equation}
  V \approx -\frac{\Delta E}{z e}.
\end{equation}

For a general magnesium intercalation compound Mg$_x$MX, the insertion reaction between two states $x_1$ and $x_2$ is given by
\begin{equation}
  \text{Mg}_{x_1}\text{MX} + (x_2 - x_1)\,\text{Mg}_{\text{metal}} \rightarrow \text{Mg}_{x_2}\text{MX}.
\end{equation}
The corresponding average voltage is computed as
\begin{equation}
  V = - \frac{E(\text{Mg}_{x_2}\text{MX}) - E(\text{Mg}_{x_1}\text{MX}) - (x_2 - x_1) E(\text{Mg}_{\text{metal}})}{z (x_2 - x_1) e},
\end{equation}
where $E(\text{Mg}_{x}\text{MX})$ are the DFT total energies of the cathode at different Mg concentrations, and $E(\text{Mg}_{\text{metal}})$ is the energy per atom of bulk magnesium. This formulation is widely used in recent computational studies of magnesium-ion batteries to evaluate voltage profiles and electrochemical performance ~\cite{han2018density,chen2025ai}.
In practice, all total energies are referenced to their fully relaxed ground-state structures.


\begin{thebibliography}{59}%
\makeatletter
\providecommand \@ifxundefined [1]{%
 \@ifx{#1\undefined}
}%
\providecommand \@ifnum [1]{%
 \ifnum #1\expandafter \@firstoftwo
 \else \expandafter \@secondoftwo
 \fi
}%
\providecommand \@ifx [1]{%
 \ifx #1\expandafter \@firstoftwo
 \else \expandafter \@secondoftwo
 \fi
}%
\providecommand \natexlab [1]{#1}%
\providecommand \enquote  [1]{``#1''}%
\providecommand \bibnamefont  [1]{#1}%
\providecommand \bibfnamefont [1]{#1}%
\providecommand \citenamefont [1]{#1}%
\providecommand \href@noop [0]{\@secondoftwo}%
\providecommand \href [0]{\begingroup \@sanitize@url \@href}%
\providecommand \@href[1]{\@@startlink{#1}\@@href}%
\providecommand \@@href[1]{\endgroup#1\@@endlink}%
\providecommand \@sanitize@url [0]{\catcode `\\12\catcode `\$12\catcode
  `\&12\catcode `\#12\catcode `\^12\catcode `\_12\catcode `\%12\relax}%
\providecommand \@@startlink[1]{}%
\providecommand \@@endlink[0]{}%
\providecommand \url  [0]{\begingroup\@sanitize@url \@url }%
\providecommand \@url [1]{\endgroup\@href {#1}{\urlprefix }}%
\providecommand \urlprefix  [0]{URL }%
\providecommand \Eprint [0]{\href }%
\providecommand \doibase [0]{https://doi.org/}%
\providecommand \selectlanguage [0]{\@gobble}%
\providecommand \bibinfo  [0]{\@secondoftwo}%
\providecommand \bibfield  [0]{\@secondoftwo}%
\providecommand \translation [1]{[#1]}%
\providecommand \BibitemOpen [0]{}%
\providecommand \bibitemStop [0]{}%
\providecommand \bibitemNoStop [0]{.\EOS\space}%
\providecommand \EOS [0]{\spacefactor3000\relax}%
\providecommand \BibitemShut  [1]{\csname bibitem#1\endcsname}%
\let\auto@bib@innerbib\@empty
\bibitem [{\citenamefont {Yoo}\ \emph {et~al.}(2013)\citenamefont {Yoo},
  \citenamefont {Shterenberg}, \citenamefont {Gofer}, \citenamefont
  {Gershinsky}, \citenamefont {Pour},\ and\ \citenamefont
  {Aurbach}}]{yoo2013mg}%
  \BibitemOpen
  \bibfield  {author} {\bibinfo {author} {\bibfnamefont {H.~D.}\ \bibnamefont
  {Yoo}}, \bibinfo {author} {\bibfnamefont {I.}~\bibnamefont {Shterenberg}},
  \bibinfo {author} {\bibfnamefont {Y.}~\bibnamefont {Gofer}}, \bibinfo
  {author} {\bibfnamefont {G.}~\bibnamefont {Gershinsky}}, \bibinfo {author}
  {\bibfnamefont {N.}~\bibnamefont {Pour}},\ and\ \bibinfo {author}
  {\bibfnamefont {D.}~\bibnamefont {Aurbach}},\ }\bibfield  {title} {\bibinfo
  {title} {Mg rechargeable batteries: an on-going challenge},\ }\href@noop {}
  {\bibfield  {journal} {\bibinfo  {journal} {Energy \& Environmental Science}\
  }\textbf {\bibinfo {volume} {6}},\ \bibinfo {pages} {2265} (\bibinfo {year}
  {2013})}\BibitemShut {NoStop}%
\bibitem [{\citenamefont {Mohtadi}\ and\ \citenamefont
  {Mizuno}(2014)}]{mohtadi2014magnesium}%
  \BibitemOpen
  \bibfield  {author} {\bibinfo {author} {\bibfnamefont {R.}~\bibnamefont
  {Mohtadi}}\ and\ \bibinfo {author} {\bibfnamefont {F.}~\bibnamefont
  {Mizuno}},\ }\bibfield  {title} {\bibinfo {title} {Magnesium batteries:
  Current state of the art, issues and future perspectives},\ }\href@noop {}
  {\bibfield  {journal} {\bibinfo  {journal} {Beilstein journal of
  nanotechnology}\ }\textbf {\bibinfo {volume} {5}},\ \bibinfo {pages} {1291}
  (\bibinfo {year} {2014})}\BibitemShut {NoStop}%
\bibitem [{\citenamefont {Chen}\ \emph {et~al.}(2021)\citenamefont {Chen},
  \citenamefont {Kang}, \citenamefont {Zhao}, \citenamefont {Wang},
  \citenamefont {Liu}, \citenamefont {Li}, \citenamefont {Liang}, \citenamefont
  {He}, \citenamefont {Li}, \citenamefont {Tavajohi},\ and\ \citenamefont
  {Baohua}}]{chen2021review}%
  \BibitemOpen
  \bibfield  {author} {\bibinfo {author} {\bibfnamefont {Y.}~\bibnamefont
  {Chen}}, \bibinfo {author} {\bibfnamefont {Y.}~\bibnamefont {Kang}}, \bibinfo
  {author} {\bibfnamefont {Y.}~\bibnamefont {Zhao}}, \bibinfo {author}
  {\bibfnamefont {L.}~\bibnamefont {Wang}}, \bibinfo {author} {\bibfnamefont
  {J.}~\bibnamefont {Liu}}, \bibinfo {author} {\bibfnamefont {Y.}~\bibnamefont
  {Li}}, \bibinfo {author} {\bibfnamefont {Z.}~\bibnamefont {Liang}}, \bibinfo
  {author} {\bibfnamefont {X.}~\bibnamefont {He}}, \bibinfo {author}
  {\bibfnamefont {X.}~\bibnamefont {Li}}, \bibinfo {author} {\bibfnamefont
  {N.}~\bibnamefont {Tavajohi}},\ and\ \bibinfo {author} {\bibfnamefont
  {L.}~\bibnamefont {Baohua}},\ }\bibfield  {title} {\bibinfo {title} {A review
  of lithium-ion battery safety concerns: The issues, strategies, and testing
  standards},\ }\href@noop {} {\bibfield  {journal} {\bibinfo  {journal}
  {Journal of Energy Chemistry}\ }\textbf {\bibinfo {volume} {59}},\ \bibinfo
  {pages} {83} (\bibinfo {year} {2021})}\BibitemShut {NoStop}%
\bibitem [{\citenamefont {Adebanjo}\ \emph {et~al.}(2025)\citenamefont
  {Adebanjo}, \citenamefont {Eko}, \citenamefont {Agbeyegbe}, \citenamefont
  {Yuk}, \citenamefont {Cowart}, \citenamefont {Nagelli}, \citenamefont
  {Burpo}, \citenamefont {Allen}, \citenamefont {Tran}, \citenamefont
  {Bhattarai}, \citenamefont {Shah}, \citenamefont {Hwang},\ and\ \citenamefont
  {Sun}}]{adebanjo2025comprehensive}%
  \BibitemOpen
  \bibfield  {author} {\bibinfo {author} {\bibfnamefont {I.~T.}\ \bibnamefont
  {Adebanjo}}, \bibinfo {author} {\bibfnamefont {J.}~\bibnamefont {Eko}},
  \bibinfo {author} {\bibfnamefont {A.~G.}\ \bibnamefont {Agbeyegbe}}, \bibinfo
  {author} {\bibfnamefont {S.~F.}\ \bibnamefont {Yuk}}, \bibinfo {author}
  {\bibfnamefont {S.~V.}\ \bibnamefont {Cowart}}, \bibinfo {author}
  {\bibfnamefont {E.~A.}\ \bibnamefont {Nagelli}}, \bibinfo {author}
  {\bibfnamefont {F.~J.}\ \bibnamefont {Burpo}}, \bibinfo {author}
  {\bibfnamefont {J.~L.}\ \bibnamefont {Allen}}, \bibinfo {author}
  {\bibfnamefont {D.~T.}\ \bibnamefont {Tran}}, \bibinfo {author}
  {\bibfnamefont {N.}~\bibnamefont {Bhattarai}}, \bibinfo {author}
  {\bibfnamefont {K.}~\bibnamefont {Shah}}, \bibinfo {author} {\bibfnamefont
  {J.-Y.}\ \bibnamefont {Hwang}},\ and\ \bibinfo {author} {\bibfnamefont
  {H.~H.}\ \bibnamefont {Sun}},\ }\bibfield  {title} {\bibinfo {title} {A
  comprehensive review of lithium-ion battery components degradation and
  operational considerations: a safety perspective},\ }\href@noop {} {\bibfield
   {journal} {\bibinfo  {journal} {Energy Advances}\ }\textbf {\bibinfo
  {volume} {4}},\ \bibinfo {pages} {820} (\bibinfo {year} {2025})}\BibitemShut
  {NoStop}%
\bibitem [{\citenamefont {Yagi}\ \emph {et~al.}(2014)\citenamefont {Yagi},
  \citenamefont {Ichitsubo}, \citenamefont {Shirai}, \citenamefont {Yanai},
  \citenamefont {Doi}, \citenamefont {Murase},\ and\ \citenamefont
  {Matsubara}}]{yagi2014concept}%
  \BibitemOpen
  \bibfield  {author} {\bibinfo {author} {\bibfnamefont {S.}~\bibnamefont
  {Yagi}}, \bibinfo {author} {\bibfnamefont {T.}~\bibnamefont {Ichitsubo}},
  \bibinfo {author} {\bibfnamefont {Y.}~\bibnamefont {Shirai}}, \bibinfo
  {author} {\bibfnamefont {S.}~\bibnamefont {Yanai}}, \bibinfo {author}
  {\bibfnamefont {T.}~\bibnamefont {Doi}}, \bibinfo {author} {\bibfnamefont
  {K.}~\bibnamefont {Murase}},\ and\ \bibinfo {author} {\bibfnamefont
  {E.}~\bibnamefont {Matsubara}},\ }\bibfield  {title} {\bibinfo {title} {A
  concept of dual-salt polyvalent-metal storage battery},\ }\href@noop {}
  {\bibfield  {journal} {\bibinfo  {journal} {Journal of Materials Chemistry
  A}\ }\textbf {\bibinfo {volume} {2}},\ \bibinfo {pages} {1144} (\bibinfo
  {year} {2014})}\BibitemShut {NoStop}%
\bibitem [{\citenamefont {Levi}\ \emph {et~al.}(2009)\citenamefont {Levi},
  \citenamefont {Levi}, \citenamefont {Chasid},\ and\ \citenamefont
  {Aurbach}}]{levi2009review}%
  \BibitemOpen
  \bibfield  {author} {\bibinfo {author} {\bibfnamefont {E.}~\bibnamefont
  {Levi}}, \bibinfo {author} {\bibfnamefont {M.}~\bibnamefont {Levi}}, \bibinfo
  {author} {\bibfnamefont {O.}~\bibnamefont {Chasid}},\ and\ \bibinfo {author}
  {\bibfnamefont {D.}~\bibnamefont {Aurbach}},\ }\bibfield  {title} {\bibinfo
  {title} {A review on the problems of the solid state ions diffusion in
  cathodes for rechargeable mg batteries},\ }\href@noop {} {\bibfield
  {journal} {\bibinfo  {journal} {Journal of Electroceramics}\ }\textbf
  {\bibinfo {volume} {22}},\ \bibinfo {pages} {13} (\bibinfo {year}
  {2009})}\BibitemShut {NoStop}%
\bibitem [{\citenamefont {Lapidus}\ \emph {et~al.}(2014)\citenamefont
  {Lapidus}, \citenamefont {Rajput}, \citenamefont {Qu}, \citenamefont
  {Chapman}, \citenamefont {Persson},\ and\ \citenamefont
  {Chupas}}]{lapidus2014solvation}%
  \BibitemOpen
  \bibfield  {author} {\bibinfo {author} {\bibfnamefont {S.~H.}\ \bibnamefont
  {Lapidus}}, \bibinfo {author} {\bibfnamefont {N.~N.}\ \bibnamefont {Rajput}},
  \bibinfo {author} {\bibfnamefont {X.}~\bibnamefont {Qu}}, \bibinfo {author}
  {\bibfnamefont {K.~W.}\ \bibnamefont {Chapman}}, \bibinfo {author}
  {\bibfnamefont {K.~A.}\ \bibnamefont {Persson}},\ and\ \bibinfo {author}
  {\bibfnamefont {P.~J.}\ \bibnamefont {Chupas}},\ }\bibfield  {title}
  {\bibinfo {title} {Solvation structure and energetics of electrolytes for
  multivalent energy storage},\ }\href@noop {} {\bibfield  {journal} {\bibinfo
  {journal} {Physical Chemistry Chemical Physics}\ }\textbf {\bibinfo {volume}
  {16}},\ \bibinfo {pages} {21941} (\bibinfo {year} {2014})}\BibitemShut
  {NoStop}%
\bibitem [{\citenamefont {Rajput}\ \emph {et~al.}(2018)\citenamefont {Rajput},
  \citenamefont {Seguin}, \citenamefont {Wood}, \citenamefont {Qu},\ and\
  \citenamefont {Persson}}]{rajput2018elucidating}%
  \BibitemOpen
  \bibfield  {author} {\bibinfo {author} {\bibfnamefont {N.~N.}\ \bibnamefont
  {Rajput}}, \bibinfo {author} {\bibfnamefont {T.~J.}\ \bibnamefont {Seguin}},
  \bibinfo {author} {\bibfnamefont {B.~M.}\ \bibnamefont {Wood}}, \bibinfo
  {author} {\bibfnamefont {X.}~\bibnamefont {Qu}},\ and\ \bibinfo {author}
  {\bibfnamefont {K.~A.}\ \bibnamefont {Persson}},\ }\bibfield  {title}
  {\bibinfo {title} {Elucidating solvation structures for rational design of
  multivalent electrolytes—a review},\ }\href@noop {} {\bibfield  {journal}
  {\bibinfo  {journal} {Modeling Electrochemical Energy Storage at the Atomic
  Scale}\ ,\ \bibinfo {pages} {79}} (\bibinfo {year} {2018})}\BibitemShut
  {NoStop}%
\bibitem [{\citenamefont {Okoshi}\ \emph {et~al.}(2013)\citenamefont {Okoshi},
  \citenamefont {Yamada}, \citenamefont {Yamada},\ and\ \citenamefont
  {Nakai}}]{okoshi2013theoretical}%
  \BibitemOpen
  \bibfield  {author} {\bibinfo {author} {\bibfnamefont {M.}~\bibnamefont
  {Okoshi}}, \bibinfo {author} {\bibfnamefont {Y.}~\bibnamefont {Yamada}},
  \bibinfo {author} {\bibfnamefont {A.}~\bibnamefont {Yamada}},\ and\ \bibinfo
  {author} {\bibfnamefont {H.}~\bibnamefont {Nakai}},\ }\bibfield  {title}
  {\bibinfo {title} {Theoretical analysis on de-solvation of lithium, sodium,
  and magnesium cations to organic electrolyte solvents},\ }\href@noop {}
  {\bibfield  {journal} {\bibinfo  {journal} {Journal of the Electrochemical
  Society}\ }\textbf {\bibinfo {volume} {160}},\ \bibinfo {pages} {A2160}
  (\bibinfo {year} {2013})}\BibitemShut {NoStop}%
\bibitem [{\citenamefont {Wan}\ \emph {et~al.}(2015)\citenamefont {Wan},
  \citenamefont {Perdue}, \citenamefont {Apblett},\ and\ \citenamefont
  {Prendergast}}]{wan2015mg}%
  \BibitemOpen
  \bibfield  {author} {\bibinfo {author} {\bibfnamefont {L.~F.}\ \bibnamefont
  {Wan}}, \bibinfo {author} {\bibfnamefont {B.~R.}\ \bibnamefont {Perdue}},
  \bibinfo {author} {\bibfnamefont {C.~A.}\ \bibnamefont {Apblett}},\ and\
  \bibinfo {author} {\bibfnamefont {D.}~\bibnamefont {Prendergast}},\
  }\bibfield  {title} {\bibinfo {title} {Mg desolvation and intercalation
  mechanism at the {Mo$_6$S$_8$} chevrel phase surface},\ }\href@noop {}
  {\bibfield  {journal} {\bibinfo  {journal} {Chemistry of Materials}\ }\textbf
  {\bibinfo {volume} {27}},\ \bibinfo {pages} {5932} (\bibinfo {year}
  {2015})}\BibitemShut {NoStop}%
\bibitem [{\citenamefont {Saha}\ \emph {et~al.}(2014)\citenamefont {Saha},
  \citenamefont {Datta}, \citenamefont {Velikokhatnyi}, \citenamefont
  {Manivannan}, \citenamefont {Alman},\ and\ \citenamefont
  {Kumta}}]{saha2014rechargeable}%
  \BibitemOpen
  \bibfield  {author} {\bibinfo {author} {\bibfnamefont {P.}~\bibnamefont
  {Saha}}, \bibinfo {author} {\bibfnamefont {M.~K.}\ \bibnamefont {Datta}},
  \bibinfo {author} {\bibfnamefont {O.~I.}\ \bibnamefont {Velikokhatnyi}},
  \bibinfo {author} {\bibfnamefont {A.}~\bibnamefont {Manivannan}}, \bibinfo
  {author} {\bibfnamefont {D.}~\bibnamefont {Alman}},\ and\ \bibinfo {author}
  {\bibfnamefont {P.~N.}\ \bibnamefont {Kumta}},\ }\bibfield  {title} {\bibinfo
  {title} {Rechargeable magnesium battery: Current status and key challenges
  for the future},\ }\href@noop {} {\bibfield  {journal} {\bibinfo  {journal}
  {Progress in Materials Science}\ }\textbf {\bibinfo {volume} {66}},\ \bibinfo
  {pages} {1} (\bibinfo {year} {2014})}\BibitemShut {NoStop}%
\bibitem [{\citenamefont {Aurbach}\ \emph {et~al.}(2000)\citenamefont
  {Aurbach}, \citenamefont {Lu}, \citenamefont {Schechter}, \citenamefont
  {Gofer}, \citenamefont {Gizbar}, \citenamefont {Turgeman}, \citenamefont
  {Cohen}, \citenamefont {Moshkovich},\ and\ \citenamefont
  {Levi}}]{aurbach2000prototype}%
  \BibitemOpen
  \bibfield  {author} {\bibinfo {author} {\bibfnamefont {D.}~\bibnamefont
  {Aurbach}}, \bibinfo {author} {\bibfnamefont {Z.}~\bibnamefont {Lu}},
  \bibinfo {author} {\bibfnamefont {A.}~\bibnamefont {Schechter}}, \bibinfo
  {author} {\bibfnamefont {Y.}~\bibnamefont {Gofer}}, \bibinfo {author}
  {\bibfnamefont {H.}~\bibnamefont {Gizbar}}, \bibinfo {author} {\bibfnamefont
  {R.}~\bibnamefont {Turgeman}}, \bibinfo {author} {\bibfnamefont
  {Y.}~\bibnamefont {Cohen}}, \bibinfo {author} {\bibfnamefont
  {M.}~\bibnamefont {Moshkovich}},\ and\ \bibinfo {author} {\bibfnamefont
  {E.}~\bibnamefont {Levi}},\ }\bibfield  {title} {\bibinfo {title} {Prototype
  systems for rechargeable magnesium batteries},\ }\href@noop {} {\bibfield
  {journal} {\bibinfo  {journal} {Nature}\ }\textbf {\bibinfo {volume} {407}},\
  \bibinfo {pages} {724} (\bibinfo {year} {2000})}\BibitemShut {NoStop}%
\bibitem [{\citenamefont {Okamoto}\ \emph {et~al.}(2015)\citenamefont
  {Okamoto}, \citenamefont {Ichitsubo}, \citenamefont {Kawaguchi},
  \citenamefont {Kumagai}, \citenamefont {Oba}, \citenamefont {Yagi},
  \citenamefont {Shimokawa}, \citenamefont {Goto}, \citenamefont {Doi},\ and\
  \citenamefont {Matsubara}}]{okamoto2015intercalation}%
  \BibitemOpen
  \bibfield  {author} {\bibinfo {author} {\bibfnamefont {S.}~\bibnamefont
  {Okamoto}}, \bibinfo {author} {\bibfnamefont {T.}~\bibnamefont {Ichitsubo}},
  \bibinfo {author} {\bibfnamefont {T.}~\bibnamefont {Kawaguchi}}, \bibinfo
  {author} {\bibfnamefont {Y.}~\bibnamefont {Kumagai}}, \bibinfo {author}
  {\bibfnamefont {F.}~\bibnamefont {Oba}}, \bibinfo {author} {\bibfnamefont
  {S.}~\bibnamefont {Yagi}}, \bibinfo {author} {\bibfnamefont {K.}~\bibnamefont
  {Shimokawa}}, \bibinfo {author} {\bibfnamefont {N.}~\bibnamefont {Goto}},
  \bibinfo {author} {\bibfnamefont {T.}~\bibnamefont {Doi}},\ and\ \bibinfo
  {author} {\bibfnamefont {E.}~\bibnamefont {Matsubara}},\ }\bibfield  {title}
  {\bibinfo {title} {Intercalation and push-out process with spinel-to-rocksalt
  transition on {Mg} insertion into spinel oxides in magnesium batteries},\
  }\href@noop {} {\bibfield  {journal} {\bibinfo  {journal} {Advanced Science}\
  }\textbf {\bibinfo {volume} {2}},\ \bibinfo {pages} {1500072} (\bibinfo
  {year} {2015})}\BibitemShut {NoStop}%
\bibitem [{\citenamefont {Zeng}\ \emph {et~al.}(2017)\citenamefont {Zeng},
  \citenamefont {Yang}, \citenamefont {Lai}, \citenamefont {Huang},
  \citenamefont {Zhang}, \citenamefont {Wang},\ and\ \citenamefont
  {Zhao}}]{zeng2017promising}%
  \BibitemOpen
  \bibfield  {author} {\bibinfo {author} {\bibfnamefont {J.}~\bibnamefont
  {Zeng}}, \bibinfo {author} {\bibfnamefont {Y.}~\bibnamefont {Yang}}, \bibinfo
  {author} {\bibfnamefont {S.}~\bibnamefont {Lai}}, \bibinfo {author}
  {\bibfnamefont {J.}~\bibnamefont {Huang}}, \bibinfo {author} {\bibfnamefont
  {Y.}~\bibnamefont {Zhang}}, \bibinfo {author} {\bibfnamefont
  {J.}~\bibnamefont {Wang}},\ and\ \bibinfo {author} {\bibfnamefont
  {J.}~\bibnamefont {Zhao}},\ }\bibfield  {title} {\bibinfo {title} {A
  promising high-voltage cathode material based on mesoporous
  {Na$_3$V$_2$(PO$_4$)$_3$/C} for rechargeable magnesium batteries},\
  }\href@noop {} {\bibfield  {journal} {\bibinfo  {journal} {Chemistry--A
  European Journal}\ }\textbf {\bibinfo {volume} {23}},\ \bibinfo {pages}
  {16898} (\bibinfo {year} {2017})}\BibitemShut {NoStop}%
\bibitem [{\citenamefont {Orikasa}\ \emph {et~al.}(2014)\citenamefont
  {Orikasa}, \citenamefont {Masese}, \citenamefont {Koyama}, \citenamefont
  {Mori}, \citenamefont {Hattori}, \citenamefont {Yamamoto}, \citenamefont
  {Okado}, \citenamefont {Huang}, \citenamefont {Minato}, \citenamefont
  {Tassel}, \citenamefont {Kim}, \citenamefont {Kobayashi}, \citenamefont
  {Abe}, \citenamefont {Kageyama},\ and\ \citenamefont
  {Uchimoto}}]{orikasa2014high}%
  \BibitemOpen
  \bibfield  {author} {\bibinfo {author} {\bibfnamefont {Y.}~\bibnamefont
  {Orikasa}}, \bibinfo {author} {\bibfnamefont {T.}~\bibnamefont {Masese}},
  \bibinfo {author} {\bibfnamefont {Y.}~\bibnamefont {Koyama}}, \bibinfo
  {author} {\bibfnamefont {T.}~\bibnamefont {Mori}}, \bibinfo {author}
  {\bibfnamefont {M.}~\bibnamefont {Hattori}}, \bibinfo {author} {\bibfnamefont
  {K.}~\bibnamefont {Yamamoto}}, \bibinfo {author} {\bibfnamefont
  {T.}~\bibnamefont {Okado}}, \bibinfo {author} {\bibfnamefont {Z.-D.}\
  \bibnamefont {Huang}}, \bibinfo {author} {\bibfnamefont {T.}~\bibnamefont
  {Minato}}, \bibinfo {author} {\bibfnamefont {C.}~\bibnamefont {Tassel}},
  \bibinfo {author} {\bibfnamefont {J.}~\bibnamefont {Kim}}, \bibinfo {author}
  {\bibfnamefont {Y.}~\bibnamefont {Kobayashi}}, \bibinfo {author}
  {\bibfnamefont {T.}~\bibnamefont {Abe}}, \bibinfo {author} {\bibfnamefont
  {H.}~\bibnamefont {Kageyama}},\ and\ \bibinfo {author} {\bibfnamefont
  {Y.}~\bibnamefont {Uchimoto}},\ }\bibfield  {title} {\bibinfo {title} {High
  energy density rechargeable magnesium battery using earth-abundant and
  non-toxic elements},\ }\href@noop {} {\bibfield  {journal} {\bibinfo
  {journal} {Scientific reports}\ }\textbf {\bibinfo {volume} {4}},\ \bibinfo
  {pages} {5622} (\bibinfo {year} {2014})}\BibitemShut {NoStop}%
\bibitem [{\citenamefont {Kaewmaraya}\ \emph {et~al.}(2014)\citenamefont
  {Kaewmaraya}, \citenamefont {Ramzan}, \citenamefont {Osorio-Guill{\'e}n},\
  and\ \citenamefont {Ahuja}}]{kaewmaraya2014electronic}%
  \BibitemOpen
  \bibfield  {author} {\bibinfo {author} {\bibfnamefont {T.}~\bibnamefont
  {Kaewmaraya}}, \bibinfo {author} {\bibfnamefont {M.}~\bibnamefont {Ramzan}},
  \bibinfo {author} {\bibfnamefont {J.}~\bibnamefont {Osorio-Guill{\'e}n}},\
  and\ \bibinfo {author} {\bibfnamefont {R.}~\bibnamefont {Ahuja}},\ }\bibfield
   {title} {\bibinfo {title} {Electronic structure and ionic diffusion of green
  battery cathode material: {Mg$_2$Mo$_6$S$_8$}},\ }\href@noop {} {\bibfield
  {journal} {\bibinfo  {journal} {Solid State Ionics}\ }\textbf {\bibinfo
  {volume} {261}},\ \bibinfo {pages} {17} (\bibinfo {year} {2014})}\BibitemShut
  {NoStop}%
\bibitem [{\citenamefont {Deng}\ \emph {et~al.}(2025)\citenamefont {Deng},
  \citenamefont {Dai}, \citenamefont {Wang}, \citenamefont {Wang},
  \citenamefont {Lu}, \citenamefont {Li}, \citenamefont {Huang}, \citenamefont
  {Chen}, \citenamefont {Huang}, \citenamefont {Gao}, \citenamefont {Luo},
  \citenamefont {Tan}, \citenamefont {Li}, \citenamefont {Wang}, \citenamefont
  {Wang},\ and\ \citenamefont {Pan}}]{deng2025advanced}%
  \BibitemOpen
  \bibfield  {author} {\bibinfo {author} {\bibfnamefont {R.}~\bibnamefont
  {Deng}}, \bibinfo {author} {\bibfnamefont {C.}~\bibnamefont {Dai}}, \bibinfo
  {author} {\bibfnamefont {Z.}~\bibnamefont {Wang}}, \bibinfo {author}
  {\bibfnamefont {Y.}~\bibnamefont {Wang}}, \bibinfo {author} {\bibfnamefont
  {G.}~\bibnamefont {Lu}}, \bibinfo {author} {\bibfnamefont {C.}~\bibnamefont
  {Li}}, \bibinfo {author} {\bibfnamefont {X.}~\bibnamefont {Huang}}, \bibinfo
  {author} {\bibfnamefont {C.}~\bibnamefont {Chen}}, \bibinfo {author}
  {\bibfnamefont {J.}~\bibnamefont {Huang}}, \bibinfo {author} {\bibfnamefont
  {Z.}~\bibnamefont {Gao}}, \bibinfo {author} {\bibfnamefont {L.}~\bibnamefont
  {Luo}}, \bibinfo {author} {\bibfnamefont {S.}~\bibnamefont {Tan}}, \bibinfo
  {author} {\bibfnamefont {H.}~\bibnamefont {Li}}, \bibinfo {author}
  {\bibfnamefont {J.}~\bibnamefont {Wang}}, \bibinfo {author} {\bibfnamefont
  {J.}~\bibnamefont {Wang}},\ and\ \bibinfo {author} {\bibfnamefont
  {F.}~\bibnamefont {Pan}},\ }\bibfield  {title} {\bibinfo {title} {Advanced
  cathode for synergistic anion-cation redox reactions in magnesium-ion
  batteries—a pathway to fast diffusion/reaction kinetics},\ }\href@noop {}
  {\bibfield  {journal} {\bibinfo  {journal} {Composites Part B: Engineering}\
  }\textbf {\bibinfo {volume} {293}},\ \bibinfo {pages} {112107} (\bibinfo
  {year} {2025})}\BibitemShut {NoStop}%
\bibitem [{\citenamefont {Saranya}\ \emph {et~al.}(2025)\citenamefont
  {Saranya}, \citenamefont {Vanitha}, \citenamefont {Sundaramahalingam},
  \citenamefont {Shameem},\ and\ \citenamefont
  {Nallamuthu}}]{saranya2025comparative}%
  \BibitemOpen
  \bibfield  {author} {\bibinfo {author} {\bibfnamefont {P.}~\bibnamefont
  {Saranya}}, \bibinfo {author} {\bibfnamefont {D.}~\bibnamefont {Vanitha}},
  \bibinfo {author} {\bibfnamefont {K.}~\bibnamefont {Sundaramahalingam}},
  \bibinfo {author} {\bibfnamefont {A.}~\bibnamefont {Shameem}},\ and\ \bibinfo
  {author} {\bibfnamefont {N.}~\bibnamefont {Nallamuthu}},\ }\bibfield  {title}
  {\bibinfo {title} {Comparative performance analysis of nanostructured metal
  oxides as cathode in solid state magnesium battery},\ }\href@noop {}
  {\bibfield  {journal} {\bibinfo  {journal} {Inorganic Chemistry
  Communications}\ }\textbf {\bibinfo {volume} {178}},\ \bibinfo {pages}
  {114458} (\bibinfo {year} {2025})}\BibitemShut {NoStop}%
\bibitem [{\citenamefont {Chen}\ \emph {et~al.}(2025)\citenamefont {Chen},
  \citenamefont {Lin}, \citenamefont {Zhang}, \citenamefont {Zhou},\ and\
  \citenamefont {Zhang}}]{chen2025ai}%
  \BibitemOpen
  \bibfield  {author} {\bibinfo {author} {\bibfnamefont {W.}~\bibnamefont
  {Chen}}, \bibinfo {author} {\bibfnamefont {Z.}~\bibnamefont {Lin}}, \bibinfo
  {author} {\bibfnamefont {X.}~\bibnamefont {Zhang}}, \bibinfo {author}
  {\bibfnamefont {H.}~\bibnamefont {Zhou}},\ and\ \bibinfo {author}
  {\bibfnamefont {Y.}~\bibnamefont {Zhang}},\ }\bibfield  {title} {\bibinfo
  {title} {Ai-driven accelerated discovery of intercalation-type cathode
  materials for magnesium batteries},\ }\href@noop {} {\bibfield  {journal}
  {\bibinfo  {journal} {Journal of Energy Chemistry}\ }\textbf {\bibinfo
  {volume} {108}},\ \bibinfo {pages} {40} (\bibinfo {year} {2025})}\BibitemShut
  {NoStop}%
\bibitem [{\citenamefont {Obeid}\ and\ \citenamefont
  {Sun}(2022)}]{obeid2022recent}%
  \BibitemOpen
  \bibfield  {author} {\bibinfo {author} {\bibfnamefont {M.~M.}\ \bibnamefont
  {Obeid}}\ and\ \bibinfo {author} {\bibfnamefont {Q.}~\bibnamefont {Sun}},\
  }\bibfield  {title} {\bibinfo {title} {Recent advances in topological quantum
  anode materials for metal-ion batteries},\ }\href@noop {} {\bibfield
  {journal} {\bibinfo  {journal} {Journal of Power Sources}\ }\textbf {\bibinfo
  {volume} {540}},\ \bibinfo {pages} {231655} (\bibinfo {year}
  {2022})}\BibitemShut {NoStop}%
\bibitem [{\citenamefont {Yi}\ \emph {et~al.}(2019)\citenamefont {Yi},
  \citenamefont {Li}, \citenamefont {Li}, \citenamefont {Zhou}, \citenamefont
  {Ma},\ and\ \citenamefont {Sun}}]{yi2019topological}%
  \BibitemOpen
  \bibfield  {author} {\bibinfo {author} {\bibfnamefont {X.}~\bibnamefont
  {Yi}}, \bibinfo {author} {\bibfnamefont {W.}~\bibnamefont {Li}}, \bibinfo
  {author} {\bibfnamefont {Z.}~\bibnamefont {Li}}, \bibinfo {author}
  {\bibfnamefont {P.}~\bibnamefont {Zhou}}, \bibinfo {author} {\bibfnamefont
  {Z.}~\bibnamefont {Ma}},\ and\ \bibinfo {author} {\bibfnamefont
  {L.}~\bibnamefont {Sun}},\ }\bibfield  {title} {\bibinfo {title} {Topological
  dual double node-line semimetals {NaAlSi (Ge)} and their potential as cathode
  material for sodium ion batteries},\ }\href@noop {} {\bibfield  {journal}
  {\bibinfo  {journal} {Journal of Materials Chemistry C}\ }\textbf {\bibinfo
  {volume} {7}},\ \bibinfo {pages} {15375} (\bibinfo {year}
  {2019})}\BibitemShut {NoStop}%
\bibitem [{\citenamefont {Wu}\ \emph {et~al.}(2020)\citenamefont {Wu},
  \citenamefont {Liang}, \citenamefont {Pang}, \citenamefont {Zhou},
  \citenamefont {Cheng}, \citenamefont {Zhang}, \citenamefont {Liu},
  \citenamefont {Johannessen},\ and\ \citenamefont {Guo}}]{wu2020coupling}%
  \BibitemOpen
  \bibfield  {author} {\bibinfo {author} {\bibfnamefont {Z.}~\bibnamefont
  {Wu}}, \bibinfo {author} {\bibfnamefont {G.}~\bibnamefont {Liang}}, \bibinfo
  {author} {\bibfnamefont {W.~K.}\ \bibnamefont {Pang}}, \bibinfo {author}
  {\bibfnamefont {T.}~\bibnamefont {Zhou}}, \bibinfo {author} {\bibfnamefont
  {Z.}~\bibnamefont {Cheng}}, \bibinfo {author} {\bibfnamefont
  {W.}~\bibnamefont {Zhang}}, \bibinfo {author} {\bibfnamefont
  {Y.}~\bibnamefont {Liu}}, \bibinfo {author} {\bibfnamefont {B.}~\bibnamefont
  {Johannessen}},\ and\ \bibinfo {author} {\bibfnamefont {Z.}~\bibnamefont
  {Guo}},\ }\bibfield  {title} {\bibinfo {title} {Coupling topological
  insulator {SnSb$_2$Te$_4$} nanodots with highly doped graphene for high-rate
  energy storage},\ }\href@noop {} {\bibfield  {journal} {\bibinfo  {journal}
  {Advanced Materials}\ }\textbf {\bibinfo {volume} {32}},\ \bibinfo {pages}
  {1905632} (\bibinfo {year} {2020})}\BibitemShut {NoStop}%
\bibitem [{\citenamefont {Liu}\ \emph {et~al.}(2017)\citenamefont {Liu},
  \citenamefont {Wang},\ and\ \citenamefont {Sun}}]{liu2017all}%
  \BibitemOpen
  \bibfield  {author} {\bibinfo {author} {\bibfnamefont {J.}~\bibnamefont
  {Liu}}, \bibinfo {author} {\bibfnamefont {S.}~\bibnamefont {Wang}},\ and\
  \bibinfo {author} {\bibfnamefont {Q.}~\bibnamefont {Sun}},\ }\bibfield
  {title} {\bibinfo {title} {All-carbon-based porous topological semimetal for
  {Li}-ion battery anode material},\ }\href@noop {} {\bibfield  {journal}
  {\bibinfo  {journal} {Proceedings of the National Academy of Sciences}\
  }\textbf {\bibinfo {volume} {114}},\ \bibinfo {pages} {651} (\bibinfo {year}
  {2017})}\BibitemShut {NoStop}%
\bibitem [{\citenamefont {Zhao}\ \emph {et~al.}(2022)\citenamefont {Zhao},
  \citenamefont {Lu}, \citenamefont {Li}, \citenamefont {Zhu}, \citenamefont
  {Meng}, \citenamefont {Li}, \citenamefont {Wang}, \citenamefont {Jiang},
  \citenamefont {Mo}, \citenamefont {Long}, \citenamefont {Gou}, \citenamefont
  {Li}, \citenamefont {Huang}, \citenamefont {Li}, \citenamefont {Ho},
  \citenamefont {Fan}, \citenamefont {Sui}, \citenamefont {Chen}, \citenamefont
  {Zhu}, \citenamefont {Liu},\ and\ \citenamefont {Zhi}}]{zhao2022few}%
  \BibitemOpen
  \bibfield  {author} {\bibinfo {author} {\bibfnamefont {Y.}~\bibnamefont
  {Zhao}}, \bibinfo {author} {\bibfnamefont {Y.}~\bibnamefont {Lu}}, \bibinfo
  {author} {\bibfnamefont {H.}~\bibnamefont {Li}}, \bibinfo {author}
  {\bibfnamefont {Y.}~\bibnamefont {Zhu}}, \bibinfo {author} {\bibfnamefont
  {Y.}~\bibnamefont {Meng}}, \bibinfo {author} {\bibfnamefont {N.}~\bibnamefont
  {Li}}, \bibinfo {author} {\bibfnamefont {D.}~\bibnamefont {Wang}}, \bibinfo
  {author} {\bibfnamefont {F.}~\bibnamefont {Jiang}}, \bibinfo {author}
  {\bibfnamefont {F.}~\bibnamefont {Mo}}, \bibinfo {author} {\bibfnamefont
  {C.}~\bibnamefont {Long}}, \bibinfo {author} {\bibfnamefont {Y.}~\bibnamefont
  {Gou}}, \bibinfo {author} {\bibfnamefont {X.}~\bibnamefont {Li}}, \bibinfo
  {author} {\bibfnamefont {Z.}~\bibnamefont {Huang}}, \bibinfo {author}
  {\bibfnamefont {Q.}~\bibnamefont {Li}}, \bibinfo {author} {\bibfnamefont
  {J.~C.}\ \bibnamefont {Ho}}, \bibinfo {author} {\bibfnamefont
  {J.}~\bibnamefont {Fan}}, \bibinfo {author} {\bibfnamefont {M.}~\bibnamefont
  {Sui}}, \bibinfo {author} {\bibfnamefont {F.}~\bibnamefont {Chen}}, \bibinfo
  {author} {\bibfnamefont {W.}~\bibnamefont {Zhu}}, \bibinfo {author}
  {\bibfnamefont {W.}~\bibnamefont {Liu}},\ and\ \bibinfo {author}
  {\bibfnamefont {C.}~\bibnamefont {Zhi}},\ }\bibfield  {title} {\bibinfo
  {title} {Few-layer bismuth selenide cathode for low-temperature
  quasi-solid-state aqueous zinc metal batteries},\ }\href@noop {} {\bibfield
  {journal} {\bibinfo  {journal} {Nature communications}\ }\textbf {\bibinfo
  {volume} {13}},\ \bibinfo {pages} {752} (\bibinfo {year} {2022})}\BibitemShut
  {NoStop}%
\bibitem [{\citenamefont {Wang}\ \emph {et~al.}(2023)\citenamefont {Wang},
  \citenamefont {Liu}, \citenamefont {Du}, \citenamefont {Sun},\ and\
  \citenamefont {Sun}}]{wang2023screening}%
  \BibitemOpen
  \bibfield  {author} {\bibinfo {author} {\bibfnamefont {Y.}~\bibnamefont
  {Wang}}, \bibinfo {author} {\bibfnamefont {J.}~\bibnamefont {Liu}}, \bibinfo
  {author} {\bibfnamefont {P.-H.}\ \bibnamefont {Du}}, \bibinfo {author}
  {\bibfnamefont {Z.}~\bibnamefont {Sun}},\ and\ \bibinfo {author}
  {\bibfnamefont {Q.}~\bibnamefont {Sun}},\ }\bibfield  {title} {\bibinfo
  {title} {Screening topological quantum cathode materials for {K}-ion
  batteries by graph neural network and first-principles calculations},\
  }\href@noop {} {\bibfield  {journal} {\bibinfo  {journal} {ACS Applied Energy
  Materials}\ }\textbf {\bibinfo {volume} {6}},\ \bibinfo {pages} {4503}
  (\bibinfo {year} {2023})}\BibitemShut {NoStop}%
\bibitem [{\citenamefont {Wu}\ and\ \citenamefont
  {Sun}(2021)}]{wu2021screening}%
  \BibitemOpen
  \bibfield  {author} {\bibinfo {author} {\bibfnamefont {W.}~\bibnamefont
  {Wu}}\ and\ \bibinfo {author} {\bibfnamefont {Q.}~\bibnamefont {Sun}},\
  }\bibfield  {title} {\bibinfo {title} {Screening topological quantum
  materials for {Na}-ion battery cathode},\ }\href@noop {} {\bibfield
  {journal} {\bibinfo  {journal} {ACS Materials Letters}\ }\textbf {\bibinfo
  {volume} {4}},\ \bibinfo {pages} {175} (\bibinfo {year} {2021})}\BibitemShut
  {NoStop}%
\bibitem [{\citenamefont {Xie}\ and\ \citenamefont
  {Grossman}(2018)}]{xie2018crystal}%
  \BibitemOpen
  \bibfield  {author} {\bibinfo {author} {\bibfnamefont {T.}~\bibnamefont
  {Xie}}\ and\ \bibinfo {author} {\bibfnamefont {J.~C.}\ \bibnamefont
  {Grossman}},\ }\bibfield  {title} {\bibinfo {title} {Crystal graph
  convolutional neural networks for an accurate and interpretable prediction of
  material properties},\ }\href@noop {} {\bibfield  {journal} {\bibinfo
  {journal} {Physical review letters}\ }\textbf {\bibinfo {volume} {120}},\
  \bibinfo {pages} {145301} (\bibinfo {year} {2018})}\BibitemShut {NoStop}%
\bibitem [{\citenamefont {Choudhary}\ and\ \citenamefont
  {DeCost}(2021)}]{choudhary2021atomistic}%
  \BibitemOpen
  \bibfield  {author} {\bibinfo {author} {\bibfnamefont {K.}~\bibnamefont
  {Choudhary}}\ and\ \bibinfo {author} {\bibfnamefont {B.}~\bibnamefont
  {DeCost}},\ }\bibfield  {title} {\bibinfo {title} {Atomistic line graph
  neural network for improved materials property predictions},\ }\href@noop {}
  {\bibfield  {journal} {\bibinfo  {journal} {npj Computational Materials}\
  }\textbf {\bibinfo {volume} {7}},\ \bibinfo {pages} {185} (\bibinfo {year}
  {2021})}\BibitemShut {NoStop}%
\bibitem [{\citenamefont {Chen}\ \emph {et~al.}(2019)\citenamefont {Chen},
  \citenamefont {Ye}, \citenamefont {Zuo}, \citenamefont {Zheng},\ and\
  \citenamefont {Ong}}]{chen2019graph}%
  \BibitemOpen
  \bibfield  {author} {\bibinfo {author} {\bibfnamefont {C.}~\bibnamefont
  {Chen}}, \bibinfo {author} {\bibfnamefont {W.}~\bibnamefont {Ye}}, \bibinfo
  {author} {\bibfnamefont {Y.}~\bibnamefont {Zuo}}, \bibinfo {author}
  {\bibfnamefont {C.}~\bibnamefont {Zheng}},\ and\ \bibinfo {author}
  {\bibfnamefont {S.~P.}\ \bibnamefont {Ong}},\ }\bibfield  {title} {\bibinfo
  {title} {Graph networks as a universal machine learning framework for
  molecules and crystals},\ }\href@noop {} {\bibfield  {journal} {\bibinfo
  {journal} {Chemistry of Materials}\ }\textbf {\bibinfo {volume} {31}},\
  \bibinfo {pages} {3564} (\bibinfo {year} {2019})}\BibitemShut {NoStop}%
\bibitem [{\citenamefont {Sch{\"u}tt}\ \emph {et~al.}(2018)\citenamefont
  {Sch{\"u}tt}, \citenamefont {Sauceda}, \citenamefont {Kindermans},
  \citenamefont {Tkatchenko},\ and\ \citenamefont
  {M{\"u}ller}}]{schutt2018schnet}%
  \BibitemOpen
  \bibfield  {author} {\bibinfo {author} {\bibfnamefont {K.~T.}\ \bibnamefont
  {Sch{\"u}tt}}, \bibinfo {author} {\bibfnamefont {H.~E.}\ \bibnamefont
  {Sauceda}}, \bibinfo {author} {\bibfnamefont {P.-J.}\ \bibnamefont
  {Kindermans}}, \bibinfo {author} {\bibfnamefont {A.}~\bibnamefont
  {Tkatchenko}},\ and\ \bibinfo {author} {\bibfnamefont {K.-R.}\ \bibnamefont
  {M{\"u}ller}},\ }\bibfield  {title} {\bibinfo {title} {Schnet--a deep
  learning architecture for molecules and materials},\ }\href@noop {}
  {\bibfield  {journal} {\bibinfo  {journal} {The Journal of chemical physics}\
  }\textbf {\bibinfo {volume} {148}} (\bibinfo {year} {2018})}\BibitemShut
  {NoStop}%
\bibitem [{\citenamefont {Dick}\ and\ \citenamefont
  {Fernandez-Serra}(2020)}]{dick2020machine}%
  \BibitemOpen
  \bibfield  {author} {\bibinfo {author} {\bibfnamefont {S.}~\bibnamefont
  {Dick}}\ and\ \bibinfo {author} {\bibfnamefont {M.}~\bibnamefont
  {Fernandez-Serra}},\ }\bibfield  {title} {\bibinfo {title} {Machine learning
  accurate exchange and correlation functionals of the electronic density},\
  }\href@noop {} {\bibfield  {journal} {\bibinfo  {journal} {Nature
  communications}\ }\textbf {\bibinfo {volume} {11}},\ \bibinfo {pages} {3509}
  (\bibinfo {year} {2020})}\BibitemShut {NoStop}%
\bibitem [{\citenamefont {Zeng}\ \emph {et~al.}(2023)\citenamefont {Zeng},
  \citenamefont {Zhang}, \citenamefont {Lu}, \citenamefont {Mo}, \citenamefont
  {Li}, \citenamefont {Chen}, \citenamefont {Rynik}, \citenamefont {Huang},
  \citenamefont {Li}, \citenamefont {Shi}, \citenamefont {Wang}, \citenamefont
  {Ye}, \citenamefont {Tuo}, \citenamefont {Yang}, \citenamefont {Ding},
  \citenamefont {Li}, \citenamefont {Tisi}, \citenamefont {Zeng}, \citenamefont
  {Bao}, \citenamefont {Xia} \emph {et~al.}}]{zeng2023deepmd}%
  \BibitemOpen
  \bibfield  {author} {\bibinfo {author} {\bibfnamefont {J.}~\bibnamefont
  {Zeng}}, \bibinfo {author} {\bibfnamefont {D.}~\bibnamefont {Zhang}},
  \bibinfo {author} {\bibfnamefont {D.}~\bibnamefont {Lu}}, \bibinfo {author}
  {\bibfnamefont {P.}~\bibnamefont {Mo}}, \bibinfo {author} {\bibfnamefont
  {Z.}~\bibnamefont {Li}}, \bibinfo {author} {\bibfnamefont {Y.}~\bibnamefont
  {Chen}}, \bibinfo {author} {\bibfnamefont {M.}~\bibnamefont {Rynik}},
  \bibinfo {author} {\bibfnamefont {L.}~\bibnamefont {Huang}}, \bibinfo
  {author} {\bibfnamefont {Z.}~\bibnamefont {Li}}, \bibinfo {author}
  {\bibfnamefont {S.}~\bibnamefont {Shi}}, \bibinfo {author} {\bibfnamefont
  {Y.}~\bibnamefont {Wang}}, \bibinfo {author} {\bibfnamefont {H.}~\bibnamefont
  {Ye}}, \bibinfo {author} {\bibfnamefont {P.}~\bibnamefont {Tuo}}, \bibinfo
  {author} {\bibfnamefont {J.}~\bibnamefont {Yang}}, \bibinfo {author}
  {\bibfnamefont {Y.}~\bibnamefont {Ding}}, \bibinfo {author} {\bibfnamefont
  {Y.}~\bibnamefont {Li}}, \bibinfo {author} {\bibfnamefont {D.}~\bibnamefont
  {Tisi}}, \bibinfo {author} {\bibfnamefont {Q.}~\bibnamefont {Zeng}}, \bibinfo
  {author} {\bibfnamefont {H.}~\bibnamefont {Bao}}, \bibinfo {author}
  {\bibfnamefont {Y.}~\bibnamefont {Xia}}, \emph {et~al.},\ }\bibfield  {title}
  {\bibinfo {title} {{DeePMD}-kit v2: A software package for deep potential
  models},\ }\href@noop {} {\bibfield  {journal} {\bibinfo  {journal} {The
  Journal of Chemical Physics}\ }\textbf {\bibinfo {volume} {159}} (\bibinfo
  {year} {2023})}\BibitemShut {NoStop}%
\bibitem [{\citenamefont {Gong}\ \emph {et~al.}(2023)\citenamefont {Gong},
  \citenamefont {Li}, \citenamefont {Zou}, \citenamefont {Xu}, \citenamefont
  {Duan},\ and\ \citenamefont {Xu}}]{gong2023general}%
  \BibitemOpen
  \bibfield  {author} {\bibinfo {author} {\bibfnamefont {X.}~\bibnamefont
  {Gong}}, \bibinfo {author} {\bibfnamefont {H.}~\bibnamefont {Li}}, \bibinfo
  {author} {\bibfnamefont {N.}~\bibnamefont {Zou}}, \bibinfo {author}
  {\bibfnamefont {R.}~\bibnamefont {Xu}}, \bibinfo {author} {\bibfnamefont
  {W.}~\bibnamefont {Duan}},\ and\ \bibinfo {author} {\bibfnamefont
  {Y.}~\bibnamefont {Xu}},\ }\bibfield  {title} {\bibinfo {title} {General
  framework for e (3)-equivariant neural network representation of density
  functional theory hamiltonian},\ }\href@noop {} {\bibfield  {journal}
  {\bibinfo  {journal} {Nature Communications}\ }\textbf {\bibinfo {volume}
  {14}},\ \bibinfo {pages} {2848} (\bibinfo {year} {2023})}\BibitemShut
  {NoStop}%
\bibitem [{\citenamefont {Jain}\ \emph {et~al.}(2013)\citenamefont {Jain},
  \citenamefont {Ong}, \citenamefont {Hautier}, \citenamefont {Chen},
  \citenamefont {Richards}, \citenamefont {Dacek}, \citenamefont {Cholia},
  \citenamefont {Gunter}, \citenamefont {Skinner}, \citenamefont {Ceder},\ and\
  \citenamefont {Persson}}]{jain2013commentary}%
  \BibitemOpen
  \bibfield  {author} {\bibinfo {author} {\bibfnamefont {A.}~\bibnamefont
  {Jain}}, \bibinfo {author} {\bibfnamefont {S.~P.}\ \bibnamefont {Ong}},
  \bibinfo {author} {\bibfnamefont {G.}~\bibnamefont {Hautier}}, \bibinfo
  {author} {\bibfnamefont {W.}~\bibnamefont {Chen}}, \bibinfo {author}
  {\bibfnamefont {W.~D.}\ \bibnamefont {Richards}}, \bibinfo {author}
  {\bibfnamefont {S.}~\bibnamefont {Dacek}}, \bibinfo {author} {\bibfnamefont
  {S.}~\bibnamefont {Cholia}}, \bibinfo {author} {\bibfnamefont
  {D.}~\bibnamefont {Gunter}}, \bibinfo {author} {\bibfnamefont
  {D.}~\bibnamefont {Skinner}}, \bibinfo {author} {\bibfnamefont
  {G.}~\bibnamefont {Ceder}},\ and\ \bibinfo {author} {\bibfnamefont {K.~A.}\
  \bibnamefont {Persson}},\ }\bibfield  {title} {\bibinfo {title} {Commentary:
  The materials project: A materials genome approach to accelerating materials
  innovation},\ }\href@noop {} {\bibfield  {journal} {\bibinfo  {journal} {APL
  materials}\ }\textbf {\bibinfo {volume} {1}} (\bibinfo {year}
  {2013})}\BibitemShut {NoStop}%
\bibitem [{\citenamefont {Hellenbrandt}(2004)}]{hellenbrandt2004inorganic}%
  \BibitemOpen
  \bibfield  {author} {\bibinfo {author} {\bibfnamefont {M.}~\bibnamefont
  {Hellenbrandt}},\ }\bibfield  {title} {\bibinfo {title} {The inorganic
  crystal structure database {(ICSD)}—present and future},\ }\href@noop {}
  {\bibfield  {journal} {\bibinfo  {journal} {Crystallography Reviews}\
  }\textbf {\bibinfo {volume} {10}},\ \bibinfo {pages} {17} (\bibinfo {year}
  {2004})}\BibitemShut {NoStop}%
\bibitem [{\citenamefont {Wang}\ \emph {et~al.}(2024)\citenamefont {Wang},
  \citenamefont {Ji}, \citenamefont {Liu}, \citenamefont {Liu}, \citenamefont
  {Zhang}, \citenamefont {Guo}, \citenamefont {Lin}, \citenamefont {Tao},
  \citenamefont {Kasemchainan}, \citenamefont {Jiang},\ and\ \citenamefont
  {Gao}}]{wang2024integrating}%
  \BibitemOpen
  \bibfield  {author} {\bibinfo {author} {\bibfnamefont {S.}~\bibnamefont
  {Wang}}, \bibinfo {author} {\bibfnamefont {Y.}~\bibnamefont {Ji}}, \bibinfo
  {author} {\bibfnamefont {J.}~\bibnamefont {Liu}}, \bibinfo {author}
  {\bibfnamefont {Z.}~\bibnamefont {Liu}}, \bibinfo {author} {\bibfnamefont
  {X.}~\bibnamefont {Zhang}}, \bibinfo {author} {\bibfnamefont
  {Y.}~\bibnamefont {Guo}}, \bibinfo {author} {\bibfnamefont {J.}~\bibnamefont
  {Lin}}, \bibinfo {author} {\bibfnamefont {J.}~\bibnamefont {Tao}}, \bibinfo
  {author} {\bibfnamefont {J.}~\bibnamefont {Kasemchainan}}, \bibinfo {author}
  {\bibfnamefont {Y.}~\bibnamefont {Jiang}},\ and\ \bibinfo {author}
  {\bibfnamefont {H.}~\bibnamefont {Gao}},\ }\bibfield  {title} {\bibinfo
  {title} {Integrating crystal structure and numerical data for predictive
  models of lithium-ion battery materials: A modified crystal graph
  convolutional neural networks approach},\ }\href@noop {} {\bibfield
  {journal} {\bibinfo  {journal} {Journal of Energy Storage}\ }\textbf
  {\bibinfo {volume} {80}},\ \bibinfo {pages} {110220} (\bibinfo {year}
  {2024})}\BibitemShut {NoStop}%
\bibitem [{\citenamefont {Zhang}\ \emph {et~al.}(2022)\citenamefont {Zhang},
  \citenamefont {Zhou}, \citenamefont {Lu},\ and\ \citenamefont
  {Shen}}]{zhang2022interpretable}%
  \BibitemOpen
  \bibfield  {author} {\bibinfo {author} {\bibfnamefont {X.}~\bibnamefont
  {Zhang}}, \bibinfo {author} {\bibfnamefont {J.}~\bibnamefont {Zhou}},
  \bibinfo {author} {\bibfnamefont {J.}~\bibnamefont {Lu}},\ and\ \bibinfo
  {author} {\bibfnamefont {L.}~\bibnamefont {Shen}},\ }\bibfield  {title}
  {\bibinfo {title} {Interpretable learning of voltage for electrode design of
  multivalent metal-ion batteries},\ }\href@noop {} {\bibfield  {journal}
  {\bibinfo  {journal} {npj Computational Materials}\ }\textbf {\bibinfo
  {volume} {8}},\ \bibinfo {pages} {175} (\bibinfo {year} {2022})}\BibitemShut
  {NoStop}%
\bibitem [{\citenamefont {Bradlyn}\ \emph {et~al.}(2017)\citenamefont
  {Bradlyn}, \citenamefont {Elcoro}, \citenamefont {Cano}, \citenamefont
  {Vergniory}, \citenamefont {Wang}, \citenamefont {Felser}, \citenamefont
  {Aroyo},\ and\ \citenamefont {Bernevig}}]{bradlyn2017topological}%
  \BibitemOpen
  \bibfield  {author} {\bibinfo {author} {\bibfnamefont {B.}~\bibnamefont
  {Bradlyn}}, \bibinfo {author} {\bibfnamefont {L.}~\bibnamefont {Elcoro}},
  \bibinfo {author} {\bibfnamefont {J.}~\bibnamefont {Cano}}, \bibinfo {author}
  {\bibfnamefont {M.~G.}\ \bibnamefont {Vergniory}}, \bibinfo {author}
  {\bibfnamefont {Z.}~\bibnamefont {Wang}}, \bibinfo {author} {\bibfnamefont
  {C.}~\bibnamefont {Felser}}, \bibinfo {author} {\bibfnamefont {M.~I.}\
  \bibnamefont {Aroyo}},\ and\ \bibinfo {author} {\bibfnamefont {B.~A.}\
  \bibnamefont {Bernevig}},\ }\bibfield  {title} {\bibinfo {title} {Topological
  quantum chemistry},\ }\href@noop {} {\bibfield  {journal} {\bibinfo
  {journal} {Nature}\ }\textbf {\bibinfo {volume} {547}},\ \bibinfo {pages}
  {298} (\bibinfo {year} {2017})}\BibitemShut {NoStop}%
\bibitem [{\citenamefont {Vergniory}\ \emph {et~al.}(2019)\citenamefont
  {Vergniory}, \citenamefont {Elcoro}, \citenamefont {Felser}, \citenamefont
  {Regnault}, \citenamefont {Bernevig},\ and\ \citenamefont
  {Wang}}]{vergniory2019complete}%
  \BibitemOpen
  \bibfield  {author} {\bibinfo {author} {\bibfnamefont {M.}~\bibnamefont
  {Vergniory}}, \bibinfo {author} {\bibfnamefont {L.}~\bibnamefont {Elcoro}},
  \bibinfo {author} {\bibfnamefont {C.}~\bibnamefont {Felser}}, \bibinfo
  {author} {\bibfnamefont {N.}~\bibnamefont {Regnault}}, \bibinfo {author}
  {\bibfnamefont {B.~A.}\ \bibnamefont {Bernevig}},\ and\ \bibinfo {author}
  {\bibfnamefont {Z.}~\bibnamefont {Wang}},\ }\bibfield  {title} {\bibinfo
  {title} {A complete catalogue of high-quality topological materials},\
  }\href@noop {} {\bibfield  {journal} {\bibinfo  {journal} {Nature}\ }\textbf
  {\bibinfo {volume} {566}},\ \bibinfo {pages} {480} (\bibinfo {year}
  {2019})}\BibitemShut {NoStop}%
\bibitem [{\citenamefont {Vergniory}\ \emph {et~al.}(2022)\citenamefont
  {Vergniory}, \citenamefont {Wieder}, \citenamefont {Elcoro}, \citenamefont
  {Parkin}, \citenamefont {Felser}, \citenamefont {Bernevig},\ and\
  \citenamefont {Regnault}}]{vergniory2022all}%
  \BibitemOpen
  \bibfield  {author} {\bibinfo {author} {\bibfnamefont {M.~G.}\ \bibnamefont
  {Vergniory}}, \bibinfo {author} {\bibfnamefont {B.~J.}\ \bibnamefont
  {Wieder}}, \bibinfo {author} {\bibfnamefont {L.}~\bibnamefont {Elcoro}},
  \bibinfo {author} {\bibfnamefont {S.~S.}\ \bibnamefont {Parkin}}, \bibinfo
  {author} {\bibfnamefont {C.}~\bibnamefont {Felser}}, \bibinfo {author}
  {\bibfnamefont {B.~A.}\ \bibnamefont {Bernevig}},\ and\ \bibinfo {author}
  {\bibfnamefont {N.}~\bibnamefont {Regnault}},\ }\bibfield  {title} {\bibinfo
  {title} {All topological bands of all nonmagnetic stoichiometric materials},\
  }\href@noop {} {\bibfield  {journal} {\bibinfo  {journal} {Science}\ }\textbf
  {\bibinfo {volume} {376}},\ \bibinfo {pages} {eabg9094} (\bibinfo {year}
  {2022})}\BibitemShut {NoStop}%
\bibitem [{\citenamefont {Urban}\ \emph {et~al.}(2016)\citenamefont {Urban},
  \citenamefont {Seo},\ and\ \citenamefont {Ceder}}]{urban2016}%
  \BibitemOpen
  \bibfield  {author} {\bibinfo {author} {\bibfnamefont {A.}~\bibnamefont
  {Urban}}, \bibinfo {author} {\bibfnamefont {D.-H.}\ \bibnamefont {Seo}},\
  and\ \bibinfo {author} {\bibfnamefont {G.}~\bibnamefont {Ceder}},\ }\bibfield
   {title} {\bibinfo {title} {Computational understanding of {Li}-ion
  batteries},\ }\href@noop {} {\bibfield  {journal} {\bibinfo  {journal} {npj
  Computational Materials}\ }\textbf {\bibinfo {volume} {2}},\ \bibinfo {pages}
  {16002} (\bibinfo {year} {2016})}\BibitemShut {NoStop}%
\bibitem [{\citenamefont {Giannozzi}\ \emph {et~al.}(2009)\citenamefont
  {Giannozzi}, \citenamefont {Baroni}, \citenamefont {Bonini}, \citenamefont
  {Calandra}, \citenamefont {Car}, \citenamefont {Cavazzoni}, \citenamefont
  {Ceresoli}, \citenamefont {Chiarotti}, \citenamefont {Cococcioni},
  \citenamefont {Dabo}, \citenamefont {Dal~Corso}, \citenamefont
  {de~Gironcoli}, \citenamefont {Fabris}, \citenamefont {Fratesi},
  \citenamefont {Gebauer}, \citenamefont {Gerstmann}, \citenamefont
  {Gougoussis}, \citenamefont {Kokalji}, \citenamefont {Lazzeri}, \citenamefont
  {Martin-Samos} \emph {et~al.}}]{giannozzi2009quantum}%
  \BibitemOpen
  \bibfield  {author} {\bibinfo {author} {\bibfnamefont {P.}~\bibnamefont
  {Giannozzi}}, \bibinfo {author} {\bibfnamefont {S.}~\bibnamefont {Baroni}},
  \bibinfo {author} {\bibfnamefont {N.}~\bibnamefont {Bonini}}, \bibinfo
  {author} {\bibfnamefont {M.}~\bibnamefont {Calandra}}, \bibinfo {author}
  {\bibfnamefont {R.}~\bibnamefont {Car}}, \bibinfo {author} {\bibfnamefont
  {C.}~\bibnamefont {Cavazzoni}}, \bibinfo {author} {\bibfnamefont
  {D.}~\bibnamefont {Ceresoli}}, \bibinfo {author} {\bibfnamefont {G.~L.}\
  \bibnamefont {Chiarotti}}, \bibinfo {author} {\bibfnamefont {M.}~\bibnamefont
  {Cococcioni}}, \bibinfo {author} {\bibfnamefont {I.}~\bibnamefont {Dabo}},
  \bibinfo {author} {\bibfnamefont {A.}~\bibnamefont {Dal~Corso}}, \bibinfo
  {author} {\bibfnamefont {S.}~\bibnamefont {de~Gironcoli}}, \bibinfo {author}
  {\bibfnamefont {S.}~\bibnamefont {Fabris}}, \bibinfo {author} {\bibfnamefont
  {G.}~\bibnamefont {Fratesi}}, \bibinfo {author} {\bibfnamefont
  {R.}~\bibnamefont {Gebauer}}, \bibinfo {author} {\bibfnamefont
  {U.}~\bibnamefont {Gerstmann}}, \bibinfo {author} {\bibfnamefont
  {C.}~\bibnamefont {Gougoussis}}, \bibinfo {author} {\bibfnamefont
  {A.}~\bibnamefont {Kokalji}}, \bibinfo {author} {\bibfnamefont
  {M.}~\bibnamefont {Lazzeri}}, \bibinfo {author} {\bibfnamefont
  {L.}~\bibnamefont {Martin-Samos}}, \emph {et~al.},\ }\bibfield  {title}
  {\bibinfo {title} {Quantum espresso: a modular and open-source software
  project for quantumsimulations of materials},\ }\href@noop {} {\bibfield
  {journal} {\bibinfo  {journal} {Journal of physics: Condensed matter}\
  }\textbf {\bibinfo {volume} {21}},\ \bibinfo {pages} {395502} (\bibinfo
  {year} {2009})}\BibitemShut {NoStop}%
\bibitem [{\citenamefont {Perdew}\ and\ \citenamefont
  {Wang}(1992)}]{perdew1992accurate}%
  \BibitemOpen
  \bibfield  {author} {\bibinfo {author} {\bibfnamefont {J.~P.}\ \bibnamefont
  {Perdew}}\ and\ \bibinfo {author} {\bibfnamefont {Y.}~\bibnamefont {Wang}},\
  }\bibfield  {title} {\bibinfo {title} {Accurate and simple analytic
  representation of the electron-gas correlation energy},\ }\href@noop {}
  {\bibfield  {journal} {\bibinfo  {journal} {Physical review B}\ }\textbf
  {\bibinfo {volume} {45}},\ \bibinfo {pages} {13244} (\bibinfo {year}
  {1992})}\BibitemShut {NoStop}%
\bibitem [{\citenamefont {Perdew}\ \emph {et~al.}(1996)\citenamefont {Perdew},
  \citenamefont {Burke},\ and\ \citenamefont
  {Ernzerhof}}]{perdew1996generalized}%
  \BibitemOpen
  \bibfield  {author} {\bibinfo {author} {\bibfnamefont {J.~P.}\ \bibnamefont
  {Perdew}}, \bibinfo {author} {\bibfnamefont {K.}~\bibnamefont {Burke}},\ and\
  \bibinfo {author} {\bibfnamefont {M.}~\bibnamefont {Ernzerhof}},\ }\bibfield
  {title} {\bibinfo {title} {Generalized gradient approximation made simple},\
  }\href@noop {} {\bibfield  {journal} {\bibinfo  {journal} {Physical review
  letters}\ }\textbf {\bibinfo {volume} {77}},\ \bibinfo {pages} {3865}
  (\bibinfo {year} {1996})}\BibitemShut {NoStop}%
\bibitem [{\citenamefont {Monkhorst}\ and\ \citenamefont
  {Pack}(1976)}]{monkhorst1976special}%
  \BibitemOpen
  \bibfield  {author} {\bibinfo {author} {\bibfnamefont {H.~J.}\ \bibnamefont
  {Monkhorst}}\ and\ \bibinfo {author} {\bibfnamefont {J.~D.}\ \bibnamefont
  {Pack}},\ }\bibfield  {title} {\bibinfo {title} {Special points for
  brillouin-zone integrations},\ }\href@noop {} {\bibfield  {journal} {\bibinfo
   {journal} {Physical review B}\ }\textbf {\bibinfo {volume} {13}},\ \bibinfo
  {pages} {5188} (\bibinfo {year} {1976})}\BibitemShut {NoStop}%
\bibitem [{\citenamefont {Zhou}\ \emph {et~al.}(2004)\citenamefont {Zhou},
  \citenamefont {Cococcioni}, \citenamefont {Marianetti}, \citenamefont
  {Morgan},\ and\ \citenamefont {Ceder}}]{zhou2004first}%
  \BibitemOpen
  \bibfield  {author} {\bibinfo {author} {\bibfnamefont {F.}~\bibnamefont
  {Zhou}}, \bibinfo {author} {\bibfnamefont {M.}~\bibnamefont {Cococcioni}},
  \bibinfo {author} {\bibfnamefont {C.~A.}\ \bibnamefont {Marianetti}},
  \bibinfo {author} {\bibfnamefont {D.}~\bibnamefont {Morgan}},\ and\ \bibinfo
  {author} {\bibfnamefont {G.}~\bibnamefont {Ceder}},\ }\bibfield  {title}
  {\bibinfo {title} {First-principles prediction of redox potentials in
  transition-metal compounds with {LDA+ U}},\ }\href@noop {} {\bibfield
  {journal} {\bibinfo  {journal} {Physical Review B—Condensed Matter and
  Materials Physics}\ }\textbf {\bibinfo {volume} {70}},\ \bibinfo {pages}
  {235121} (\bibinfo {year} {2004})}\BibitemShut {NoStop}%
\bibitem [{\citenamefont {Cococcioni}\ and\ \citenamefont
  {De~Gironcoli}(2005)}]{cococcioni2005linear}%
  \BibitemOpen
  \bibfield  {author} {\bibinfo {author} {\bibfnamefont {M.}~\bibnamefont
  {Cococcioni}}\ and\ \bibinfo {author} {\bibfnamefont {S.}~\bibnamefont
  {De~Gironcoli}},\ }\bibfield  {title} {\bibinfo {title} {Linear response
  approach to the calculation of the effective interaction parameters in the
  lda+ u method},\ }\href@noop {} {\bibfield  {journal} {\bibinfo  {journal}
  {Physical Review B—Condensed Matter and Materials Physics}\ }\textbf
  {\bibinfo {volume} {71}},\ \bibinfo {pages} {035105} (\bibinfo {year}
  {2005})}\BibitemShut {NoStop}%
\bibitem [{\citenamefont {Wang}\ \emph {et~al.}(2006)\citenamefont {Wang},
  \citenamefont {Maxisch},\ and\ \citenamefont {Ceder}}]{wang2006oxidation}%
  \BibitemOpen
  \bibfield  {author} {\bibinfo {author} {\bibfnamefont {L.}~\bibnamefont
  {Wang}}, \bibinfo {author} {\bibfnamefont {T.}~\bibnamefont {Maxisch}},\ and\
  \bibinfo {author} {\bibfnamefont {G.}~\bibnamefont {Ceder}},\ }\bibfield
  {title} {\bibinfo {title} {Oxidation energies of transition metal oxides
  within the {GGA+ U} framework},\ }\href@noop {} {\bibfield  {journal}
  {\bibinfo  {journal} {Physical Review B—Condensed Matter and Materials
  Physics}\ }\textbf {\bibinfo {volume} {73}},\ \bibinfo {pages} {195107}
  (\bibinfo {year} {2006})}\BibitemShut {NoStop}%
\bibitem [{\citenamefont {Jain}\ \emph {et~al.}(2011)\citenamefont {Jain},
  \citenamefont {Hautier}, \citenamefont {Ong}, \citenamefont {Moore},
  \citenamefont {Fischer}, \citenamefont {Persson},\ and\ \citenamefont
  {Ceder}}]{jain2011formation}%
  \BibitemOpen
  \bibfield  {author} {\bibinfo {author} {\bibfnamefont {A.}~\bibnamefont
  {Jain}}, \bibinfo {author} {\bibfnamefont {G.}~\bibnamefont {Hautier}},
  \bibinfo {author} {\bibfnamefont {S.~P.}\ \bibnamefont {Ong}}, \bibinfo
  {author} {\bibfnamefont {C.~J.}\ \bibnamefont {Moore}}, \bibinfo {author}
  {\bibfnamefont {C.~C.}\ \bibnamefont {Fischer}}, \bibinfo {author}
  {\bibfnamefont {K.~A.}\ \bibnamefont {Persson}},\ and\ \bibinfo {author}
  {\bibfnamefont {G.}~\bibnamefont {Ceder}},\ }\bibfield  {title} {\bibinfo
  {title} {Formation enthalpies by mixing {GGA} and {GGA+ U} calculations},\
  }\href@noop {} {\bibfield  {journal} {\bibinfo  {journal} {Physical Review
  B—Condensed Matter and Materials Physics}\ }\textbf {\bibinfo {volume}
  {84}},\ \bibinfo {pages} {045115} (\bibinfo {year} {2011})}\BibitemShut
  {NoStop}%
\bibitem [{\citenamefont {Munro}\ \emph {et~al.}(2020)\citenamefont {Munro},
  \citenamefont {Latimer}, \citenamefont {Horton}, \citenamefont {Dwaraknath},\
  and\ \citenamefont {Persson}}]{munro2020improved}%
  \BibitemOpen
  \bibfield  {author} {\bibinfo {author} {\bibfnamefont {J.~M.}\ \bibnamefont
  {Munro}}, \bibinfo {author} {\bibfnamefont {K.}~\bibnamefont {Latimer}},
  \bibinfo {author} {\bibfnamefont {M.~K.}\ \bibnamefont {Horton}}, \bibinfo
  {author} {\bibfnamefont {S.}~\bibnamefont {Dwaraknath}},\ and\ \bibinfo
  {author} {\bibfnamefont {K.~A.}\ \bibnamefont {Persson}},\ }\bibfield
  {title} {\bibinfo {title} {An improved symmetry-based approach to reciprocal
  space path selection in band structure calculations},\ }\href@noop {}
  {\bibfield  {journal} {\bibinfo  {journal} {npj Computational Materials}\
  }\textbf {\bibinfo {volume} {6}},\ \bibinfo {pages} {112} (\bibinfo {year}
  {2020})}\BibitemShut {NoStop}%
\bibitem [{\citenamefont {Ebrahimzadeh}\ \emph {et~al.}(2025)\citenamefont
  {Ebrahimzadeh}, \citenamefont {Sharif},\ and\ \citenamefont
  {Banad}}]{ebrahimzadeh2025accelerated}%
  \BibitemOpen
  \bibfield  {author} {\bibinfo {author} {\bibfnamefont {D.}~\bibnamefont
  {Ebrahimzadeh}}, \bibinfo {author} {\bibfnamefont {S.~S.}\ \bibnamefont
  {Sharif}},\ and\ \bibinfo {author} {\bibfnamefont {Y.~M.}\ \bibnamefont
  {Banad}},\ }\bibfield  {title} {\bibinfo {title} {Accelerated discovery of
  vanadium oxide compositions: A {WGAN-VAE} framework for materials design},\
  }\href@noop {} {\bibfield  {journal} {\bibinfo  {journal} {Materials Today
  Electronics}\ }\textbf {\bibinfo {volume} {13}},\ \bibinfo {pages} {100155}
  (\bibinfo {year} {2025})}\BibitemShut {NoStop}%
\bibitem [{\citenamefont {Wang}\ and\ \citenamefont
  {Navrotsky}(2004)}]{wang2004enthalpy}%
  \BibitemOpen
  \bibfield  {author} {\bibinfo {author} {\bibfnamefont {M.}~\bibnamefont
  {Wang}}\ and\ \bibinfo {author} {\bibfnamefont {A.}~\bibnamefont
  {Navrotsky}},\ }\bibfield  {title} {\bibinfo {title} {Enthalpy of formation
  of {LiNiO$_2$}, {LiCoO$_2$} and their solid solution, {LiNi$_1-
  x$Co$_x$O$_2$}},\ }\href@noop {} {\bibfield  {journal} {\bibinfo  {journal}
  {Solid State Ionics}\ }\textbf {\bibinfo {volume} {166}},\ \bibinfo {pages}
  {167} (\bibinfo {year} {2004})}\BibitemShut {NoStop}%
\bibitem [{\citenamefont {Basnet}\ \emph {et~al.}(2022)\citenamefont {Basnet},
  \citenamefont {Kotur}, \citenamefont {Rybak}, \citenamefont {Stephenson},
  \citenamefont {Bishop}, \citenamefont {Autieri}, \citenamefont {Birowska},\
  and\ \citenamefont {Hu}}]{basnet2022controlling}%
  \BibitemOpen
  \bibfield  {author} {\bibinfo {author} {\bibfnamefont {R.}~\bibnamefont
  {Basnet}}, \bibinfo {author} {\bibfnamefont {K.~M.}\ \bibnamefont {Kotur}},
  \bibinfo {author} {\bibfnamefont {M.}~\bibnamefont {Rybak}}, \bibinfo
  {author} {\bibfnamefont {C.}~\bibnamefont {Stephenson}}, \bibinfo {author}
  {\bibfnamefont {S.}~\bibnamefont {Bishop}}, \bibinfo {author} {\bibfnamefont
  {C.}~\bibnamefont {Autieri}}, \bibinfo {author} {\bibfnamefont
  {M.}~\bibnamefont {Birowska}},\ and\ \bibinfo {author} {\bibfnamefont
  {J.}~\bibnamefont {Hu}},\ }\bibfield  {title} {\bibinfo {title} {Controlling
  magnetic exchange and anisotropy by nonmagnetic ligand substitution in
  layered {MPX 3 (M= Ni, Mn; X= S, Se)}},\ }\href@noop {} {\bibfield  {journal}
  {\bibinfo  {journal} {Physical Review Research}\ }\textbf {\bibinfo {volume}
  {4}},\ \bibinfo {pages} {023256} (\bibinfo {year} {2022})}\BibitemShut
  {NoStop}%
\bibitem [{\citenamefont {Sun}\ \emph {et~al.}(2016)\citenamefont {Sun},
  \citenamefont {Dacek}, \citenamefont {Ong}, \citenamefont {Hautier},
  \citenamefont {Jain}, \citenamefont {Richards}, \citenamefont {Gamst},
  \citenamefont {Persson},\ and\ \citenamefont {Ceder}}]{sun2016thermodynamic}%
  \BibitemOpen
  \bibfield  {author} {\bibinfo {author} {\bibfnamefont {W.}~\bibnamefont
  {Sun}}, \bibinfo {author} {\bibfnamefont {S.~T.}\ \bibnamefont {Dacek}},
  \bibinfo {author} {\bibfnamefont {S.~P.}\ \bibnamefont {Ong}}, \bibinfo
  {author} {\bibfnamefont {G.}~\bibnamefont {Hautier}}, \bibinfo {author}
  {\bibfnamefont {A.}~\bibnamefont {Jain}}, \bibinfo {author} {\bibfnamefont
  {W.~D.}\ \bibnamefont {Richards}}, \bibinfo {author} {\bibfnamefont {A.~C.}\
  \bibnamefont {Gamst}}, \bibinfo {author} {\bibfnamefont {K.~A.}\ \bibnamefont
  {Persson}},\ and\ \bibinfo {author} {\bibfnamefont {G.}~\bibnamefont
  {Ceder}},\ }\bibfield  {title} {\bibinfo {title} {The thermodynamic scale of
  inorganic crystalline metastability},\ }\href@noop {} {\bibfield  {journal}
  {\bibinfo  {journal} {Science advances}\ }\textbf {\bibinfo {volume} {2}},\
  \bibinfo {pages} {e1600225} (\bibinfo {year} {2016})}\BibitemShut {NoStop}%
\bibitem [{\citenamefont {Han}\ \emph {et~al.}(2018)\citenamefont {Han},
  \citenamefont {Liu}, \citenamefont {Sun}, \citenamefont {Sendek},\ and\
  \citenamefont {Yang}}]{han2018density}%
  \BibitemOpen
  \bibfield  {author} {\bibinfo {author} {\bibfnamefont {X.}~\bibnamefont
  {Han}}, \bibinfo {author} {\bibfnamefont {C.}~\bibnamefont {Liu}}, \bibinfo
  {author} {\bibfnamefont {J.}~\bibnamefont {Sun}}, \bibinfo {author}
  {\bibfnamefont {A.~D.}\ \bibnamefont {Sendek}},\ and\ \bibinfo {author}
  {\bibfnamefont {W.}~\bibnamefont {Yang}},\ }\bibfield  {title} {\bibinfo
  {title} {Density functional theory calculations for evaluation of phosphorene
  as a potential anode material for magnesium batteries},\ }\href@noop {}
  {\bibfield  {journal} {\bibinfo  {journal} {RSC advances}\ }\textbf {\bibinfo
  {volume} {8}},\ \bibinfo {pages} {7196} (\bibinfo {year} {2018})}\BibitemShut
  {NoStop}%
\bibitem [{\citenamefont {Chen}\ \emph {et~al.}(2018)\citenamefont {Chen},
  \citenamefont {Sai~Gautam}, \citenamefont {Huang},\ and\ \citenamefont
  {Ceder}}]{chen2018first}%
  \BibitemOpen
  \bibfield  {author} {\bibinfo {author} {\bibfnamefont {T.}~\bibnamefont
  {Chen}}, \bibinfo {author} {\bibfnamefont {G.}~\bibnamefont {Sai~Gautam}},
  \bibinfo {author} {\bibfnamefont {W.}~\bibnamefont {Huang}},\ and\ \bibinfo
  {author} {\bibfnamefont {G.}~\bibnamefont {Ceder}},\ }\bibfield  {title}
  {\bibinfo {title} {First-principles study of the voltage profile and mobility
  of {Mg} intercalation in a chromium oxide spinel},\ }\href@noop {} {\bibfield
   {journal} {\bibinfo  {journal} {Chemistry of Materials}\ }\textbf {\bibinfo
  {volume} {30}},\ \bibinfo {pages} {153} (\bibinfo {year} {2018})}\BibitemShut
  {NoStop}%
\bibitem [{\citenamefont {Truong}\ \emph {et~al.}(2017)\citenamefont {Truong},
  \citenamefont {Kempaiah~Devaraju}, \citenamefont {Tran}, \citenamefont
  {Gambe}, \citenamefont {Nayuki}, \citenamefont {Sasaki},\ and\ \citenamefont
  {Honma}}]{truong2017unravelling}%
  \BibitemOpen
  \bibfield  {author} {\bibinfo {author} {\bibfnamefont {Q.~D.}\ \bibnamefont
  {Truong}}, \bibinfo {author} {\bibfnamefont {M.}~\bibnamefont
  {Kempaiah~Devaraju}}, \bibinfo {author} {\bibfnamefont {P.~D.}\ \bibnamefont
  {Tran}}, \bibinfo {author} {\bibfnamefont {Y.}~\bibnamefont {Gambe}},
  \bibinfo {author} {\bibfnamefont {K.}~\bibnamefont {Nayuki}}, \bibinfo
  {author} {\bibfnamefont {Y.}~\bibnamefont {Sasaki}},\ and\ \bibinfo {author}
  {\bibfnamefont {I.}~\bibnamefont {Honma}},\ }\bibfield  {title} {\bibinfo
  {title} {Unravelling the surface structure of {MgMn$_2$O$_4$} cathode
  materials for rechargeable magnesium-ion battery},\ }\href@noop {} {\bibfield
   {journal} {\bibinfo  {journal} {Chemistry of Materials}\ }\textbf {\bibinfo
  {volume} {29}},\ \bibinfo {pages} {6245} (\bibinfo {year}
  {2017})}\BibitemShut {NoStop}%
\bibitem [{\citenamefont {Mukherjee}\ \emph {et~al.}(2020)\citenamefont
  {Mukherjee}, \citenamefont {Taragin}, \citenamefont {Aviv}, \citenamefont
  {Perelshtein},\ and\ \citenamefont {Noked}}]{mukherjee2020}%
  \BibitemOpen
  \bibfield  {author} {\bibinfo {author} {\bibfnamefont {A.}~\bibnamefont
  {Mukherjee}}, \bibinfo {author} {\bibfnamefont {S.}~\bibnamefont {Taragin}},
  \bibinfo {author} {\bibfnamefont {H.}~\bibnamefont {Aviv}}, \bibinfo {author}
  {\bibfnamefont {I.}~\bibnamefont {Perelshtein}},\ and\ \bibinfo {author}
  {\bibfnamefont {M.}~\bibnamefont {Noked}},\ }\bibfield  {title} {\bibinfo
  {title} {{Rationally Designed Vanadium Pentoxide as High Capacity Insertion
  Material for Mg-Ion}},\ }\href@noop {} {\bibfield  {journal} {\bibinfo
  {journal} {Advanced Functional Materials}\ }\textbf {\bibinfo {volume}
  {30}},\ \bibinfo {pages} {2003518} (\bibinfo {year} {2020})}\BibitemShut
  {NoStop}%
\bibitem [{\citenamefont {Huie}\ \emph {et~al.}(2015)\citenamefont {Huie},
  \citenamefont {Bock}, \citenamefont {Takeuchi}, \citenamefont {Marschilok},\
  and\ \citenamefont {Takeuchi}}]{huie2015cathode}%
  \BibitemOpen
  \bibfield  {author} {\bibinfo {author} {\bibfnamefont {M.~M.}\ \bibnamefont
  {Huie}}, \bibinfo {author} {\bibfnamefont {D.~C.}\ \bibnamefont {Bock}},
  \bibinfo {author} {\bibfnamefont {E.~S.}\ \bibnamefont {Takeuchi}}, \bibinfo
  {author} {\bibfnamefont {A.~C.}\ \bibnamefont {Marschilok}},\ and\ \bibinfo
  {author} {\bibfnamefont {K.~J.}\ \bibnamefont {Takeuchi}},\ }\bibfield
  {title} {\bibinfo {title} {Cathode materials for magnesium and magnesium-ion
  based batteries},\ }\href@noop {} {\bibfield  {journal} {\bibinfo  {journal}
  {Coordination Chemistry Reviews}\ }\textbf {\bibinfo {volume} {287}},\
  \bibinfo {pages} {15} (\bibinfo {year} {2015})}\BibitemShut {NoStop}%
\end{thebibliography}
\end{document}